\documentclass{emulateapj}
\usepackage{graphicx}
\pdfoutput=1
\usepackage{color}

\submitted{ApJ in press; Received April 3, 2014; Accepted: August 19, 2014}

\def\gs{\mathrel{\raise0.35ex\hbox{$\scriptstyle >$}\kern-0.6em
\lower0.40ex\hbox{{$\scriptstyle \sim$}}}}
\def\ls{\mathrel{\raise0.35ex\hbox{$\scriptstyle <$}\kern-0.6em
\lower0.40ex\hbox{{$\scriptstyle \sim$}}}}
\newcommand\ffh{\mbox{$ \!\!^{\mathrm h}$}}%
\newcommand\ffm{\mbox{$ \!\!^{\mathrm m}$}}%
\newcommand\ffs{\mbox{$ \!\!^{\mathrm s}$}}%

\shortauthors{Owen et al.}
\shorttitle{VLA Observations of A2256 I}

\begin{document}

\title{Wideband VLA Observations of Abell 2256 I: Continuum, 
Rotation Measure and Spectral Imaging}

\author{
Frazer\,N.\ Owen,\altaffilmark{1},
Lawrence Rudnick,\altaffilmark{2},
Jean Eilek, \altaffilmark{3,4},
Urvashi Rau,\altaffilmark{1},
Sanjay Bhatnagar, \altaffilmark{1}
Leonid Kogan, \altaffilmark{1}}

\altaffiltext{1}{National Radio Astronomy Observatory, P.\ O.\ Box O,
Socorro, NM 87801 USA.; fowen@nrao.edu. The National Radio Astronomy
Observatory is facility of the National Science Foundation operated
under cooperative agreement by Associated Universities Inc.}
\altaffiltext{2}{University of Minnesota}
\altaffiltext{3} {Adjunct Astronomer at the National
Radio Astronomy Observatory} 
\altaffiltext{4}{New Mexico Tech}

\setcounter{footnote}{5}

\begin{abstract}

We report new observations of Abell 2256 with the  Karl G.
Jansky Very Large Array (VLA) at frequencies
between 1 and 8 GHz. These observations take advantage of the 2:1
bandwidths available during a single observation to study the spectral
index, polarization and Rotation Measure as well as using the
associated higher sensitivity per unit time to image total intensity 
features down to
$\sim 0.5$\arcsec\ resolution. We find the Large Relic, which
dominates the cluster,
is made up of a complex of filaments which show correlated distributions 
in intensity, spectral index, and fractional polarization. The
Rotation Measure varies across the face of the Large Relic but is not
well correlated with the other properties of the source. The shape of 
individual filaments suggests that the Large Relic is at least 25 kpc
thick.  We
detect a low surface brightness arc connecting the Large
Relic to the Halo and other radio structures suggesting a physical 
connection between these features. The center of the
F-complex is dominated by a very steep-spectrum, polarized, ring-like
structure, F2, without an obvious
optical identification, but the entire F-complex does have interesting
morphological similarities to the radio structure of NGC1265. 
Source C, the Long Tail, is
unresolved in width near the galaxy core and is $\ls 100$pc in
diameter there. This morphology suggests either that C is a one-sided 
jet or that the bending of
the tails takes place very near the core, consistent with the parent galaxy 
having undergone extreme stripping.
Overall it seems that many of the unusual phenomena can be
understood in the context of Abell 2256 
being near the pericenter of a slightly off-axis merger between a
cluster and a
smaller group. Given the lack of evidence for a strong shock
associated with the Large Relic, other models should be considered,
such as reconnection between two large-scale magnetic
domains.

\end{abstract}


\section{Introduction}

Abell 2256 contains perhaps the richest variety of radio phenomena of any
known rich cluster \citep[e.g.,][]{b76,b79,r94,c06,k10,v12a}.  The Mpc-scale,
relatively flat spectrum, diffuse structure -- often called a radio relic -- is
perhaps the most intriguing \citep[e.g.,][]{c06}. It shows many similarities 
to the general class of radio relics, which are the large, often elongated structures 
found near the periphery of many clusters \citep[e.g.,][]{f12}. Because of their 
location and the lack of any association with a cluster galaxy,  such relics are 
generally thought to be caused by large-scale shocks generated in cluster mergers
\citep[e.g.,][]{bj}.

There is also a radio
halo roughly coincident with the X-ray emission \citep[e.g.,][]{c06},
several radio tails including one which is extremely straight over its
$\gs 500$ kpc length \citep[e.g.,][]{r94,m03,b08}, a complex of very steep spectrum 
emission (the F-complex) most of which is not
clearly associated with any cluster member \citep[e.g.,][]{r94,m03}
and more than 40 cluster members with detected radio emission \citep{m03}.

 Abell 2256 ($z=0.0583$) is an Abell richness class 2, massive cluster, with an estimated total 
mass of $\sim10^{15}$ M$_\odot$  within 1 Mpc of the cluster center \citep{b02,m99}.
The cluster has been proposed to be an ongoing merger \citep{f89,f91} of
two or three
previously independent clusters \citep{b02}. The cluster appears to be
in the early or mid-stages of its merger \citep{r95,s02} and thus may allow
us to study some of the environmental changes resulting from such
mergers while they are taking place. 

The EVLA project, which has increased the frequency coverage, bandwidth
and number of channels that can be observed in a single
observation with VLA, gives us an important new tool to study these
phenomena. Here we report initial observations using these
capabilities to study the total intensity and polarization of 
the emission from Abell 2256 in the 1-8 GHz range. This paper is the
first in a series analyzing the results of these new observations. 
This paper is intended to introduce and summarize the new results we have obtained. 
Later papers are planned which will address in more detail 
1) the cluster magnetic field based on the
Faraday Rotation of individual sources, 2) the physics of the
individual radio galaxies in Abell 2256, 3) the physics of the Large
Relic, and 4) the radio properties of the cluster galaxy population.

This paper is divided into 5 sections. In \S 2  we cover
observations, editing, calibration and imaging of the VLA wideband data.
In section \S 3, we present the basic results from the project. In
section \S 4 we discuss the broader implications for understanding
Abell 2256 and in \S 5 we summarize our most important conclusions.
We will assume $H_0$=70 km s$^{-1}$ Mpc$^{-1}$, $\Omega_M=0.27$, $\Omega_{vac}=0.73$ 
in what follows.
\begin{deluxetable}{lrrrr}
\tablecolumns{4}
\tablewidth{0pt}
\tablecaption{Summary of Observations \label{obs}}
\tablenum{1}
\pagestyle{empty}
\tablehead{
\colhead{Config} & 
\colhead{Date} &
\colhead{GHz} &
\colhead{Hours}}
\startdata
D&29-JUL-2010&$1-2$&3&\\ 
C&31-OCT-2010&$1-2$&6&\\
B&21-MAR-2011&$1-2$&5&\\
B&10-APR-2011&$1-2$&6&\\
A&26-OCT-2012&$1-2$&3&\\
A&28-OCT-2012&$1-2$&3&\\
A&30-OCT-2012&$1-2$&3&\\
A&17-OCT-2012&$2-4$&3&\\
A&23-OCT-2012&$2-4$&3&\\
A&12-OCT-2012&$4-6$&3&\\
A&24-OCT-2012&$4-6$&3&\\
A&25-OCT-2012&$4-6$&3&\\
A&09-OCT-2012&$6-8$&3&\\
\enddata
\end{deluxetable}

\section{Observations, Editing, Calibration \& Imaging}

Data were obtained in the A, B, C and D configurations as summarized in
Table~\ref{obs}. The 8 or 16 subbands of 128 MHz each were used to
cover the frequency range from 1-2, 2-4, 4-6, and 6-8 GHz. 
Each
subband had 64 2MHz channels. About one third of the full bandwidth was
lost due to interference in the 1-2 GHz range, especially in the range
1500-1648 MHz. Much less, but significant interference was encountered
in the other bands.  At each band observations were made in 4 to 8 hour
scheduling blocks with the Abell 2256 field observed in scans of $20-25$ 
minutes each bracketed
by a phase calibrator, either J1634+6245 or J1800+7828. 3C48 was
used as the flux density calibrator and 3C138 was used as the polarization
position angle calibrator.

The calibration was carried out in the standard way  using AIPS, except for a few
changes needed to deal with the very wide bandwidth. The bandpass was
calculated using the phase calibrator, only without any calibration
except for the delay correction. These data were then
self-calibrated in phase only and then a single bandpass solution for
each observed track was  derived using BPASS. This procedure has the
effect of averaging out most of the interference since the interfering
signals are mostly either from satellites or from sources on the
ground and thus their observed phases change rapidly with respect to
sources in the sky. 

\begin{figure*}[htb]
\vspace{-1.2in}
\includegraphics[width=2.1\columnwidth]
{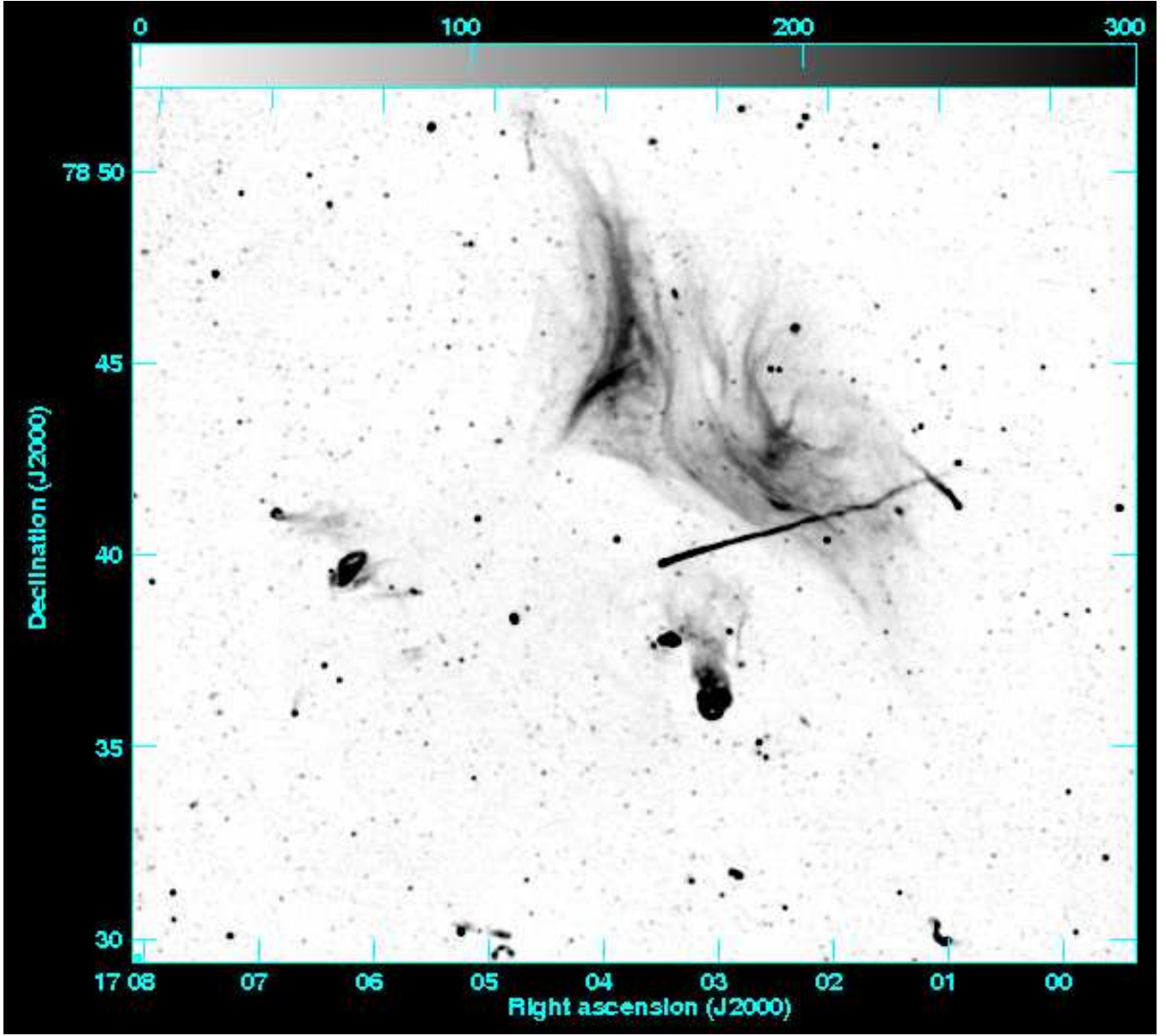}
\vspace{-1.55in}
\caption{Grey-scale radio image of A2256 at 6\arcsec\ (clean beam)
  resolution. The intensity wedge at the top of the figure shows the brightness in $\mu$Jy/beam.
\label{bwI8}} 
\end{figure*}

Using the delay and bandpass corrections only, the
data were then flagged using the AIPS task RFLAG. This task 
uses the fact that the phase of the interference changes rapidly with
respect to the astronomical sources, as well as the fact that the
interference typically is not circularly polarized and thus shows
up much more prominently in the cross-hand correlations used to
observe linear polarization. RFLAG was used to calculate the vector
rms of groups of three adjacent integrations (in time) on each baseline in each
channel in each subband for the RL and LR correlations. Then 
all the correlations were flagged for the middle of the three
integrations when the rms was found to be about $5\sigma$ above the mean rms 
for that subband and baseline. For more details see \citet{aips}.
Flagging was also done for the first
two and last two channels for each subband and a few other channels that
were spoiled by the correlator. The 1500-1648 and 2128-2384 MHz
frequency ranges were also entirely
removed. This procedure cleans up enough of the interference so that
it is not a serious problem for the rest of the calibration and
imaging.

However the sensitivity does still vary across the full band and with
respect to time due to the instrument itself and to the residual
interference. After amplitude and phase calibration, the task REWAY
was used to calibrate the weights by calculating the rms as a function
of time for each baseline and subband across the unflagged
channels. In this way an empirical weight could be assigned to each
visibility in calibrated units of Jy$^{-2}$.

The ionospheric Rotation Measure (RM) was
corrected using the AIPS procedure VLATECR which uses a time variable
atmospheric model of the ionosphere and the earth's magnetic field. In
practice these corrections were negligible during all the times of
these observations. The polarization calibration was done in the
standard way in AIPS \citep{aips} except that
the calibration for instrumental polarization and position angle were
done as a function of channel.  A narrowband image of the Abell 2256
field near the center of each band was made
in AIPS using the task IMAGR. These images were used to self-calibrate the 
phases for each band before applying the polarization corrections. 
For each subband the position angle
calibration was generally accurate to better than 1 degree and the instrumental
polarization was corrected to better than a few tenths of a percent.


\begin{figure*}
\vspace{-0.7in}
\includegraphics[width=1.15\columnwidth]
{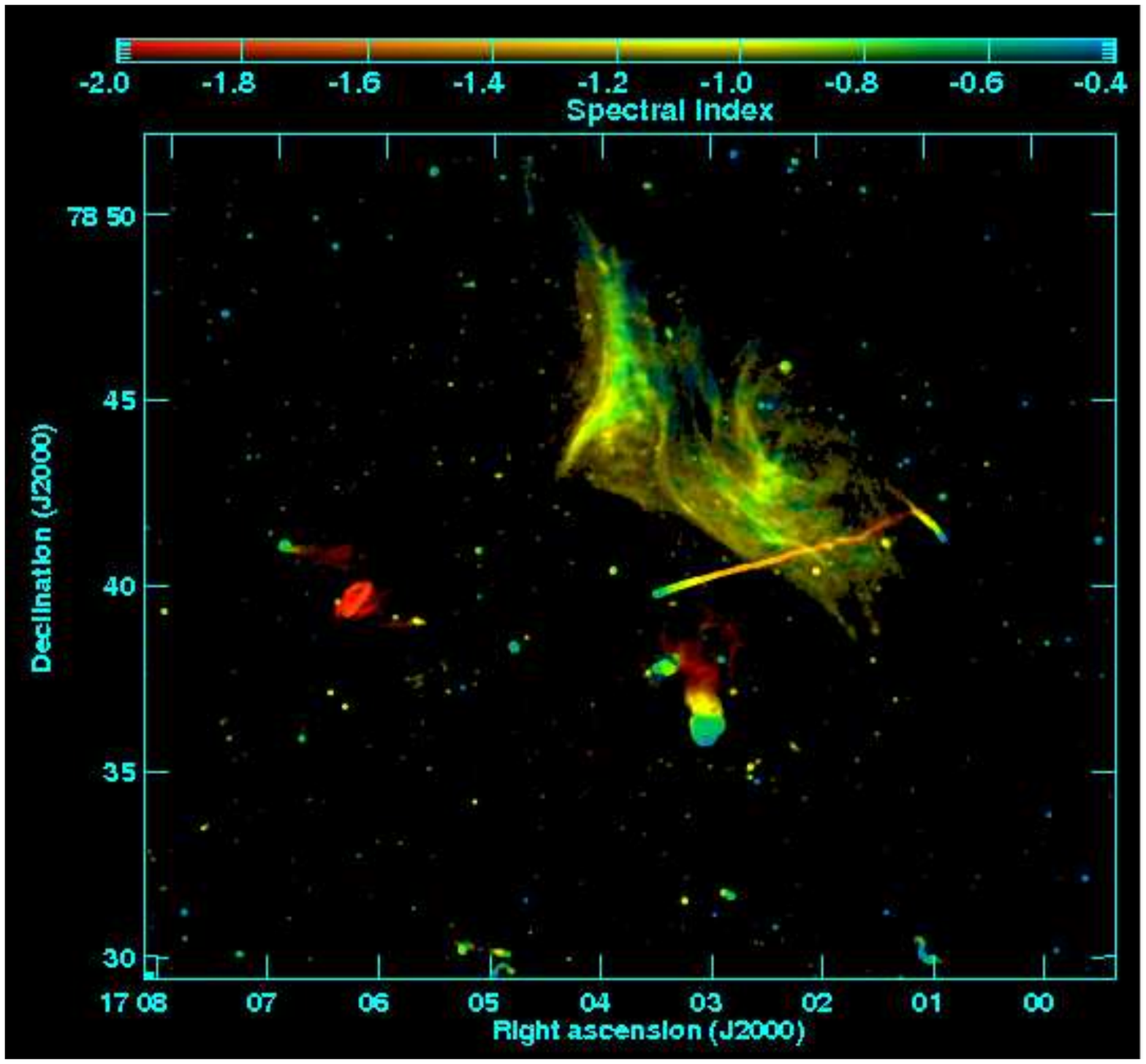}
\vspace{-0.1\columnwidth}
\hspace{-0.075\columnwidth}
\includegraphics[width=1.15\columnwidth]
{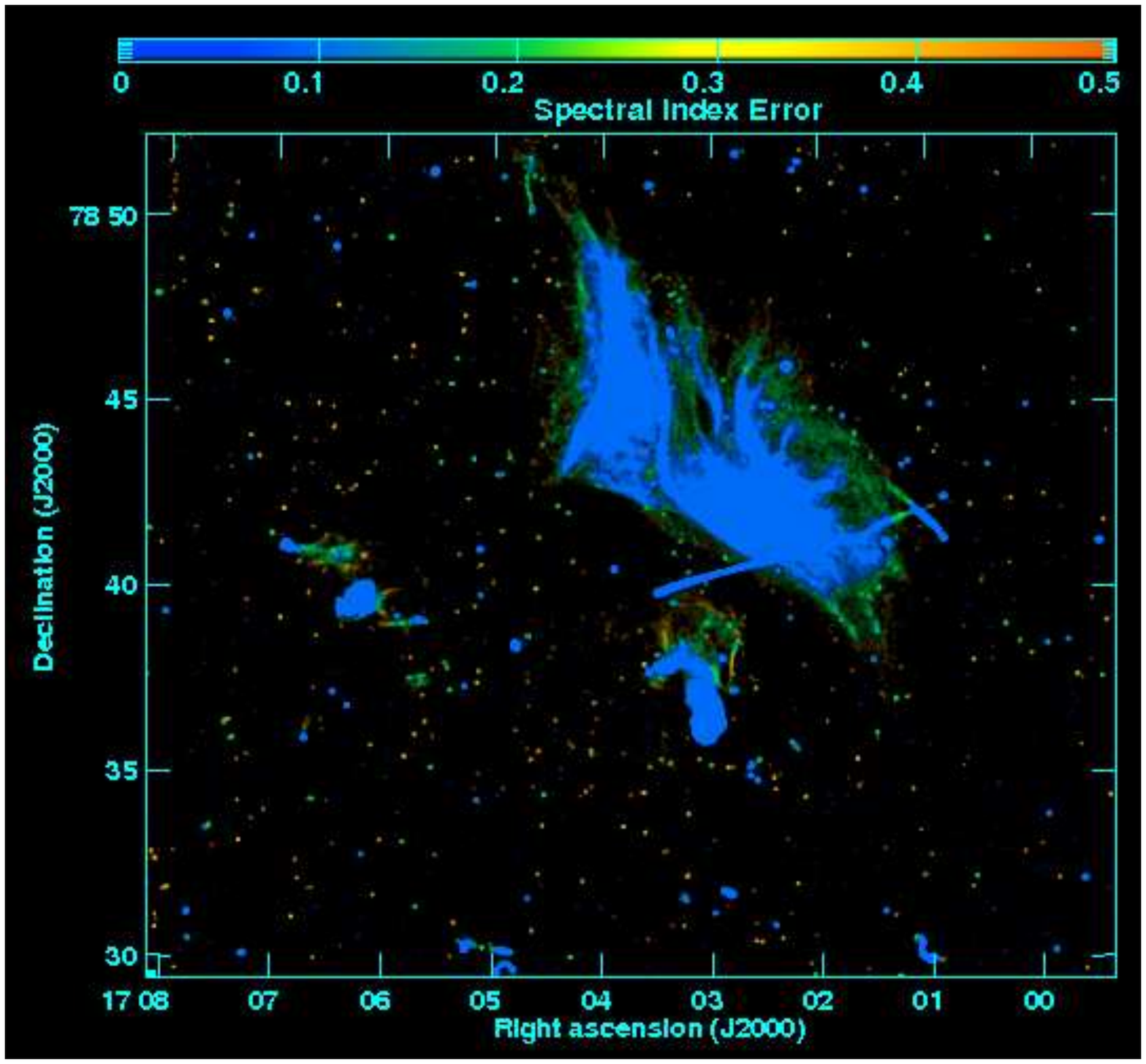}
\vspace{0.02\columnwidth}
\vspace{-1.65in}
\\
\includegraphics[width=1.15\columnwidth]
{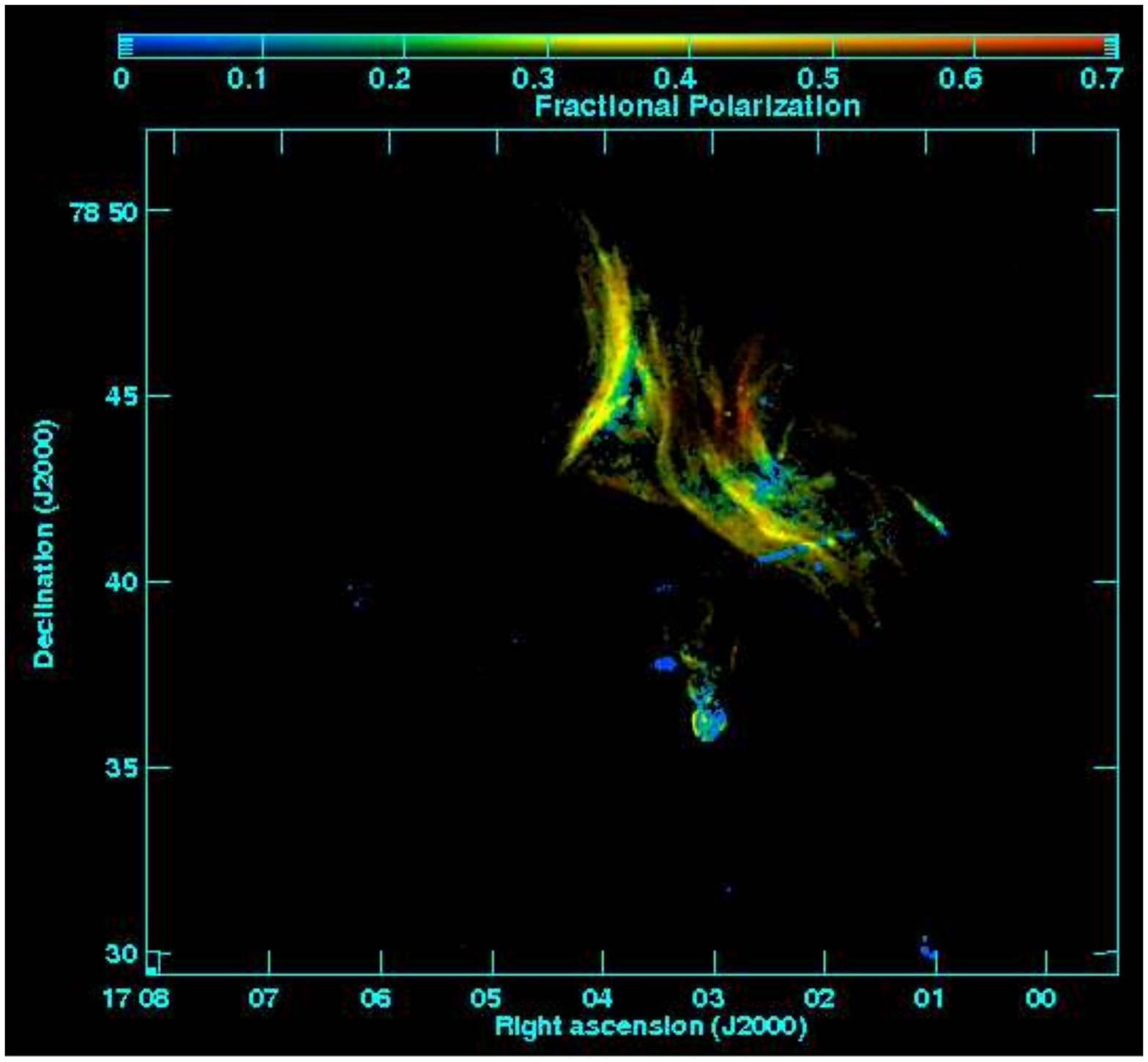}
\hspace{-0.07\columnwidth}
\includegraphics[width=1.15\columnwidth]
{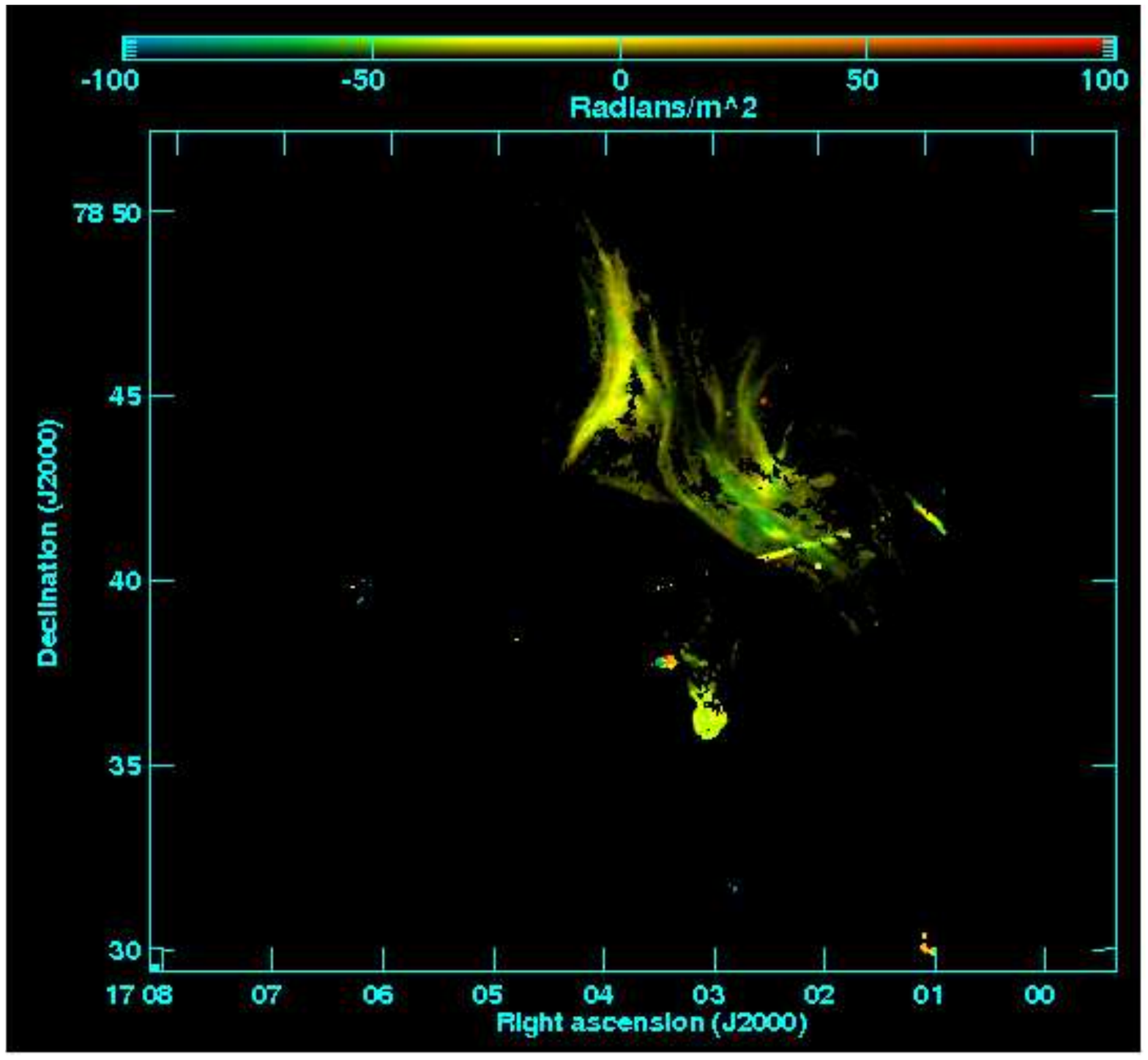}
\vspace{-1.0in}
\caption{Wide-field images of the Abell 2256 field.  Top left,
``true'' color radio image of A2256. The intensity image
has a 6\arcsec\ circular clean beam and the color has a 12\arcsec\ circular clean
beam.The color bar shows spectral 
index $\alpha$ from -2.0 to -0.4, $S\propto \nu^{\alpha}$. Top right,
spectral index error image of A2256  quantized in
  steps,$0-0.1,0.1-0.2,0.2-0.3,0.3-0.4$ and $0.4-0.5$  with the same clean beam sizes as
the top left panel.  Lower left, Fractional Polarization at 
6\arcsec\ resolution. The color bar shows the fractional polarization
levels from 0 to 0.70.  Lower right, RM Max from AFARS at 6\arcsec\
resolution. The color bar shows RM values between $-100$ and $+100$
rad/m$^2$. RM values beyond the -100 to 
100 rad/m$^2$  range are plotted as bright blue ($< -100$) or bright red ($> +100$).
\label{fourway_1}}
\end{figure*}

\begin{figure}[htb]
\vspace{0.22in}
\includegraphics[width=1.0\columnwidth]
{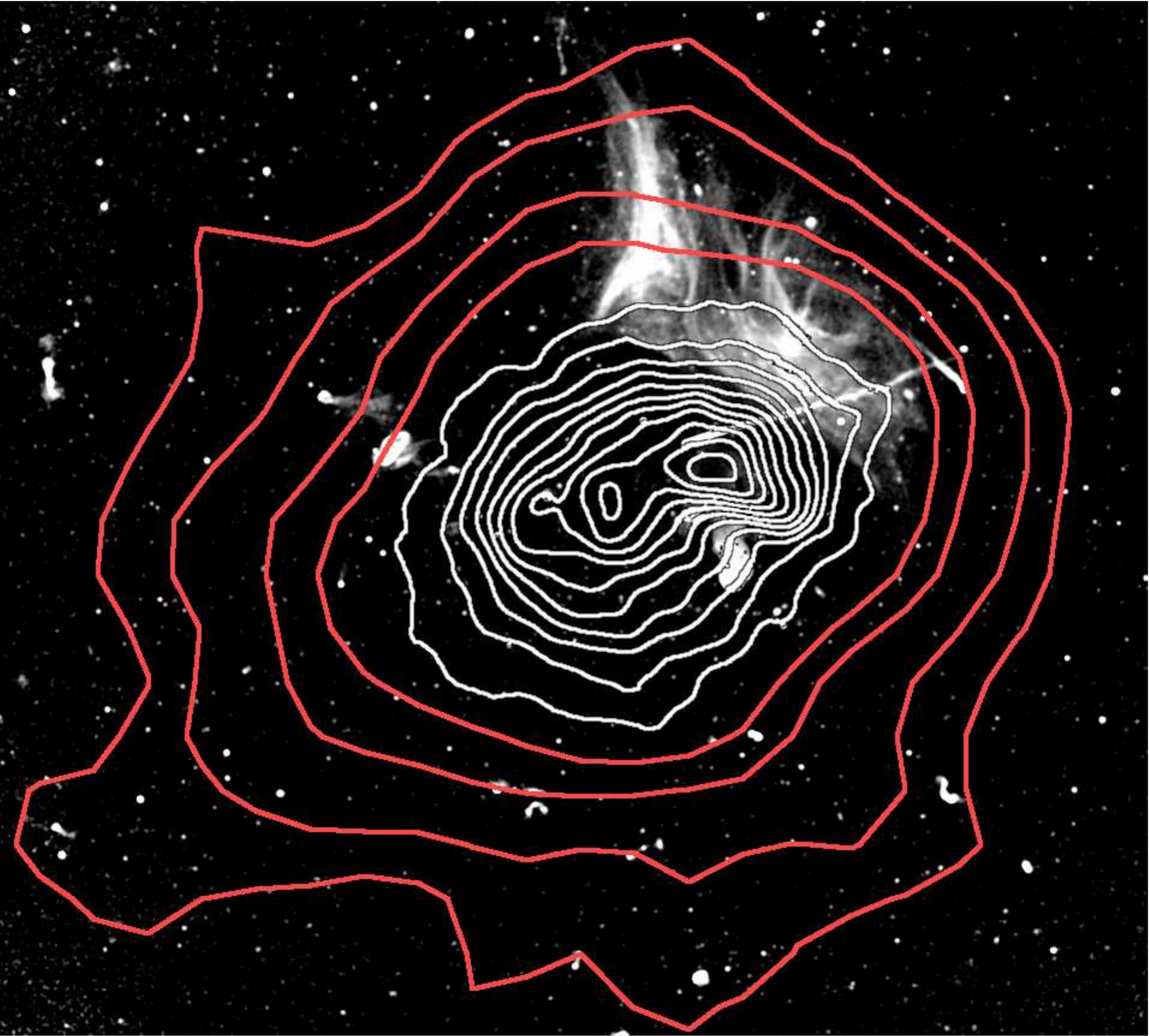}
\vspace{-0.02in}
\caption{Low resolution grey scale image with 12\arcsec\ resolution
  with X-ray overlay contours. The inner (white) contours are from
  {\it Chandra} and the outer ({ red} contours are from ROSAT.
\label{BX}} 
\end{figure}

The total intensity imaging was carried out in CASA using the 
MSMFS deconvolution algorithm \citep{r11}. This algorithm, as implemented
in the CASA task {\tt clean}, deconvolves the image using multiple scales
and multiple Taylor coefficients as a function of frequency to describe
the spatial intensity distribution over the full observed bandwidth. CASA clean uses the
W-projection which corrects for the sky curvature, so that we could
make one large image which covered the
entire field.  The output products of interest
are a total intensity image at a reference frequency and a spectral
index image.  After imaging with CASA
clean, the CASA task {\tt widebandpbcor} was used to correct all the output
images for the primary beam attenuation. 

In addition we provide spectral index error images.
The errors in the spectral index image are a combination of random
errors in the calibrated visibilities and imperfections in the
deconvolution process.
In order to include both error sources, the spectral index error map
is derived by propagating residual errors in the individual Taylor 
coefficient maps through the division that calculates the spectral index value.
To account for per pixel residuals as well as global noise levels, an error estimate was
computed as

\begin{equation}
\Delta I_{\alpha}  =  \sqrt{  \left(\Delta I_{\alpha}^{global}\right)^2  +  \left( g~\Delta I_{\alpha}^{pixel} \right)^2}
\end{equation}

\noindent where $ \Delta I_{\alpha}^{global} $ and $\Delta
I_{\alpha}^{pixel}$ are calculated according to equation 39 of
\citet{r11}, with the global error per Taylor coefficient computed 
as the median of the absolute deviation from the median of all pixel 
amplitudes in the residual image, and the per-pixel error read off 
directly from the Taylor coefficient residual images. We determined the
value of $g$ from the small-scale spectral index scatter in the 6\arcsec\ resolution
spectral index image. We find the best value for $g$ is $0.03$. In
this way we have generated a spectral index error estimate at each
point in the image. The same value for $g$, 0.03, is used for the
12\arcsec\ image which we find is consistent with the scatter in
$\alpha$ on that image.

The polarization imaging was done in AIPS. We first calculated cubes
of Q and U using 10 MHz channels. The spectral range below 1250 MHz
was not useful due to the final polarizers not being available for
L-band (1-2 GHz). The AIPS task IMAGR was used with multi-scales and
faceting appropriate to correct for the sky curvature. The facets were
assembled into two  monolithic cubes using FLATN which cover the
entire field. 

The output cubes were corrected for spectral index and primary beam
attenuation with the task SPCOR, using the spectral index image from
CASA. Then the task FARS was used to calculate a RM-synthesis cube \citep{b05} from
which we can derive, among other things the maximum polarized flux and
the corresponding Rotation Measure for each pixel. These techniques
will be discussed more in a later paper. 

The total intensity, polarization and RM-synthesis images are each
calculated at a number of resolutions to emphasize different features
in the data. Each image was made using all of the unflagged data.
Various combinations of Briggs robust weighting in CASA and a uv-taper were
used to modify the output synthesized beam. A range of modeling scales, ranging from
0 to the maximum minor axis size of a feature bright enough to be detected, 
were used in the MSMFS algorithm for each image in order to cover all
the accessible size 
scales. To obtain circular beams for the total intensity images
in CASA, we used a Gaussian smoothing script on the MSMFS images with somewhat 
smaller clean beams, in a way which is consistent with the 
MSMFS formalism. For the RM-synthesis imaging, we made  each narrowband Q and U image
in AIPS with a taper calculated so that the clean beam was slightly smaller than the
desired circular clean beam and then convolved the images, using CONVL, to the desired circular
clean beam size.  A subset of these will be discussed below.

\section{Results}

\subsection{Total Intensity and Spectral Index Images}

In figure~\ref{bwI8}, we show the total intensity image of the
center of the Abell 2256 field with a 6\arcsec\ circular
restoring beam. This image and the many others that follow, 
especially the four panel figures, have more details than can easily be seen in
the letter-size pages of the printed journal. We encourage the
reader who is viewing the online or pdf versions of this paper to
blow up these figures to examine the finer-scale features.
In figures~\ref{fourway_1} top left and
\ref{fourway_1} top right, we display the
corresponding ``true''\footnote{By ``true'' color we mean that
  the color represents the measured spectral index ranging from
  emission relatively brighter at longer wavelengths (red) to emission
  brighter at shorter wavelengths (blue) as the human eye might see them if
  it were sensitive to radio wavelengths.} color image showing the spectral indices and
their errors respectively. These images show the primary results of
our total intensity imaging: the heavily filamented Large Relic; the
Long Tail with a gradually steepening spectral index distribution; the
central complex south of the Long Tail; the steep spectrum
(red) complex to the east of the of the main concentration, consisting
of a narrow angle tail source identified with a cluster member 
and a ring of emission not obviously attached to any cluster member;
and a variety of other cluster member radio sources.  

Much of the discussion of the physics of A2256 depends on the relation of
the radio emission to the X-ray emission. For this perspective, we display in
figure~\ref{BX} X-ray contours made from a combination of the ROSAT
(outer, red contours) \citep{b91} and the {\it Chandra} (inner,
white contours) \citep{s02} images overlaid on the 12\arcsec\
resolution, grey scale radio image.

\subsection{Polarization \& RM Synthesis}

The polarization images shown in this paper were made using the
AIPS task, AFARS, which at each pixel searches the FARS RM-synthesis 
cube for the maximum amplitude and outputs two RA/Dec images, one of
the maximum amplitude and one of the corresponding RM. If the
RM is due to a simple foreground screen then these two images are
the total polarized flux density for each pixel and the RM from that
screen. For this paper we will only present these results and other
images derived from them using the total intensity images. In later
papers we will perform a more detailed analysis.

In figures~\ref{fourway_1} bottom left and~\ref{fourway_1} bottom
right, we show the
fractional polarization and the peak RM  for the 
same region as for the total intensity image in figure~\ref{bwI8}. For this 
RM synthesis using FARS, a range of RM from $-200$ to $+200$ rad
m$^{-2}$ was searched with
a RM resolution of 1 rad m$^{-2}$. In practice little interesting RM
structure is seen outside of the range $-100$ to $100$ rad m$^{-2}$,
so only that range is displayed  in order to show some of the RM structure
without hiding the larger range of RMs visible in the field. Later in this paper we
show more detail for the Large Relic and further papers will display and 
discuss the RM structure in detail for the other sources.
For both images the polarization intensity has been corrected for
noise bias and
has been clipped at $20\mu$Jy, about $4\sigma$ above the noise in the AFARS maximum 
amplitude image. Fainter features can be seen at lower S/N along with a larger number of
spurious detections. At higher resolution and with a more limited search range in RM, 
some of the additional structure becomes more significant.

\subsection{Large Scale Features}

\begin{figure}
\vspace{-0.7in}
\includegraphics[width=1.0\columnwidth]
{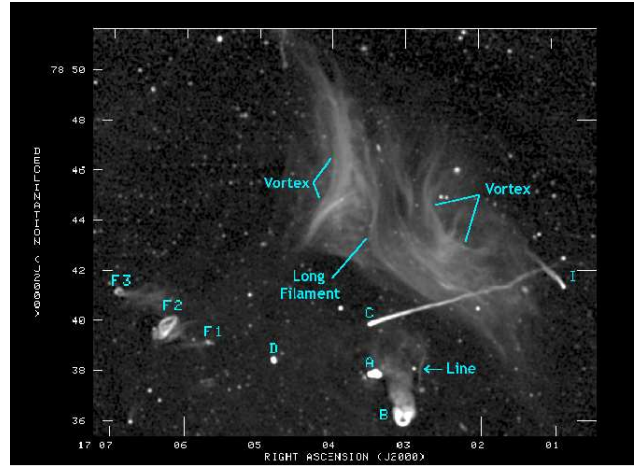}
\vspace{-1.1in}
\caption{Grey-scale radio image of A2256 at 6\arcsec\ resolution with source labels.
\label{bw6l}} 
\end{figure}

\begin{figure*}
\vspace{-0.7in}
\includegraphics[width=1.16\columnwidth,angle=0,trim=0.0cm 0.0cm 0.0cm 0.0cm,clip=True]
{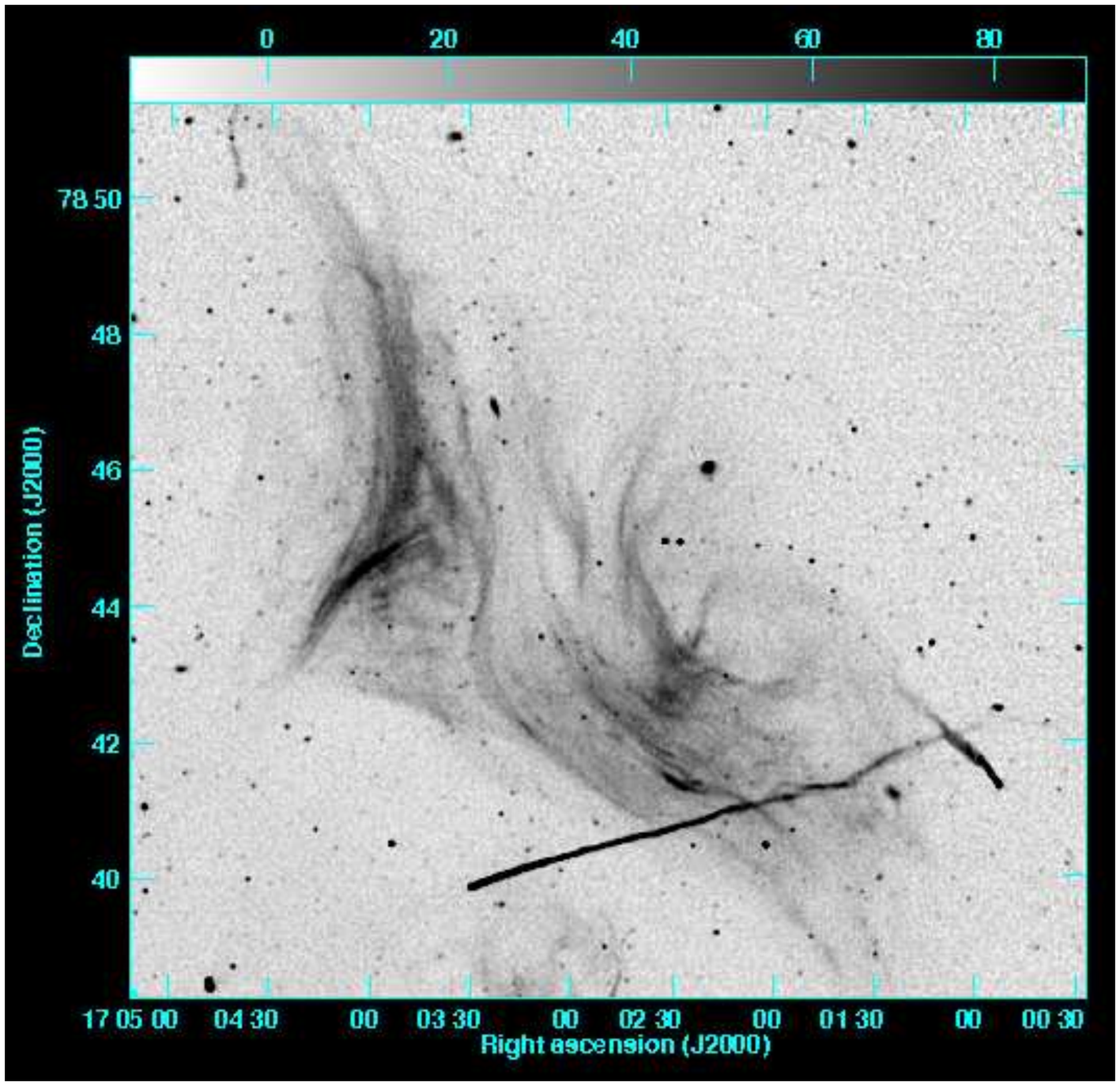}
\hspace{-0.14\columnwidth}
\includegraphics[width=1.15\columnwidth,angle=0,trim=0.0cm 0cm 0cm 0cm,clip=True]
{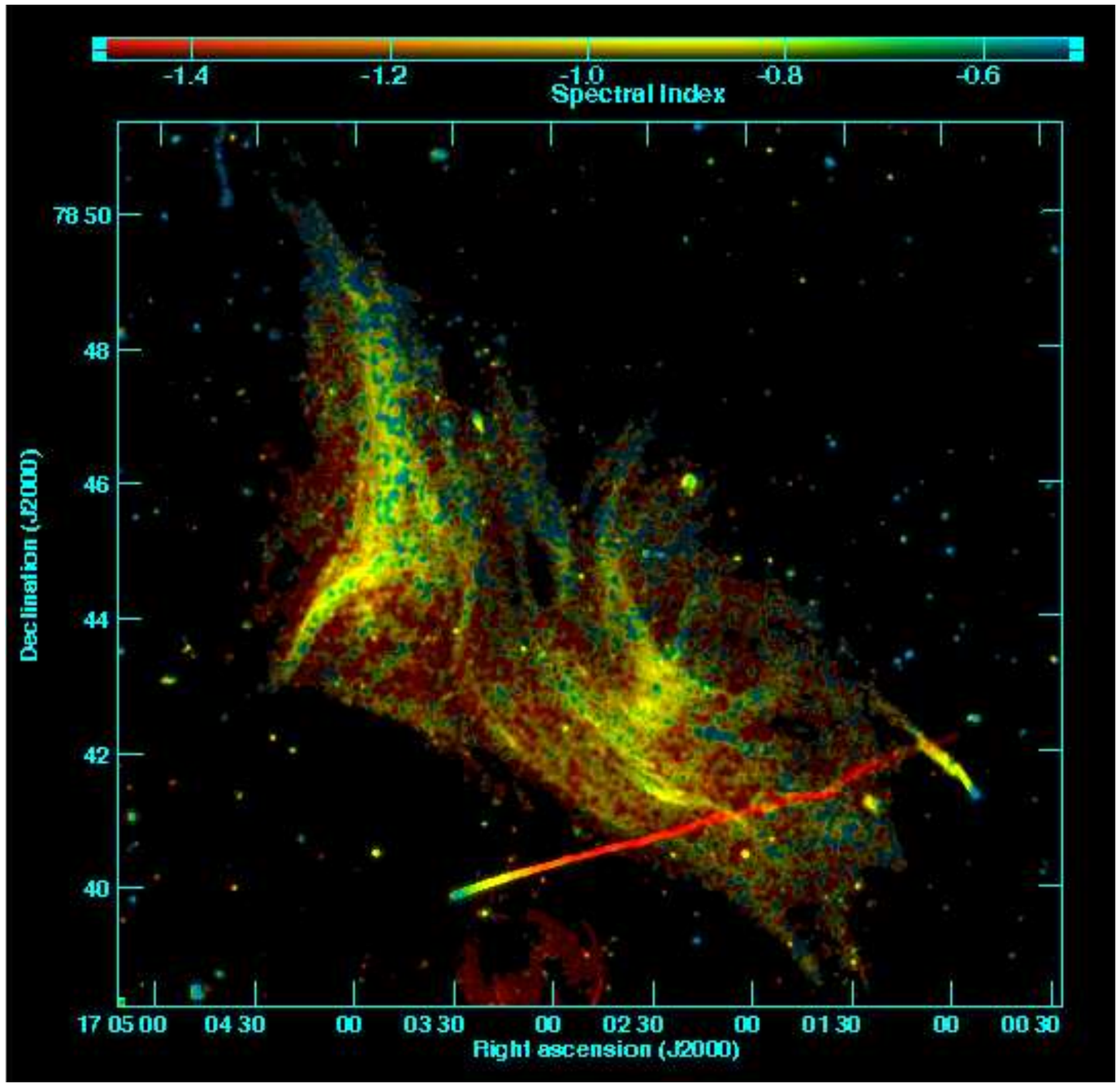}
\vspace{-1.6in}
\\
\includegraphics[width=1.15\columnwidth]
{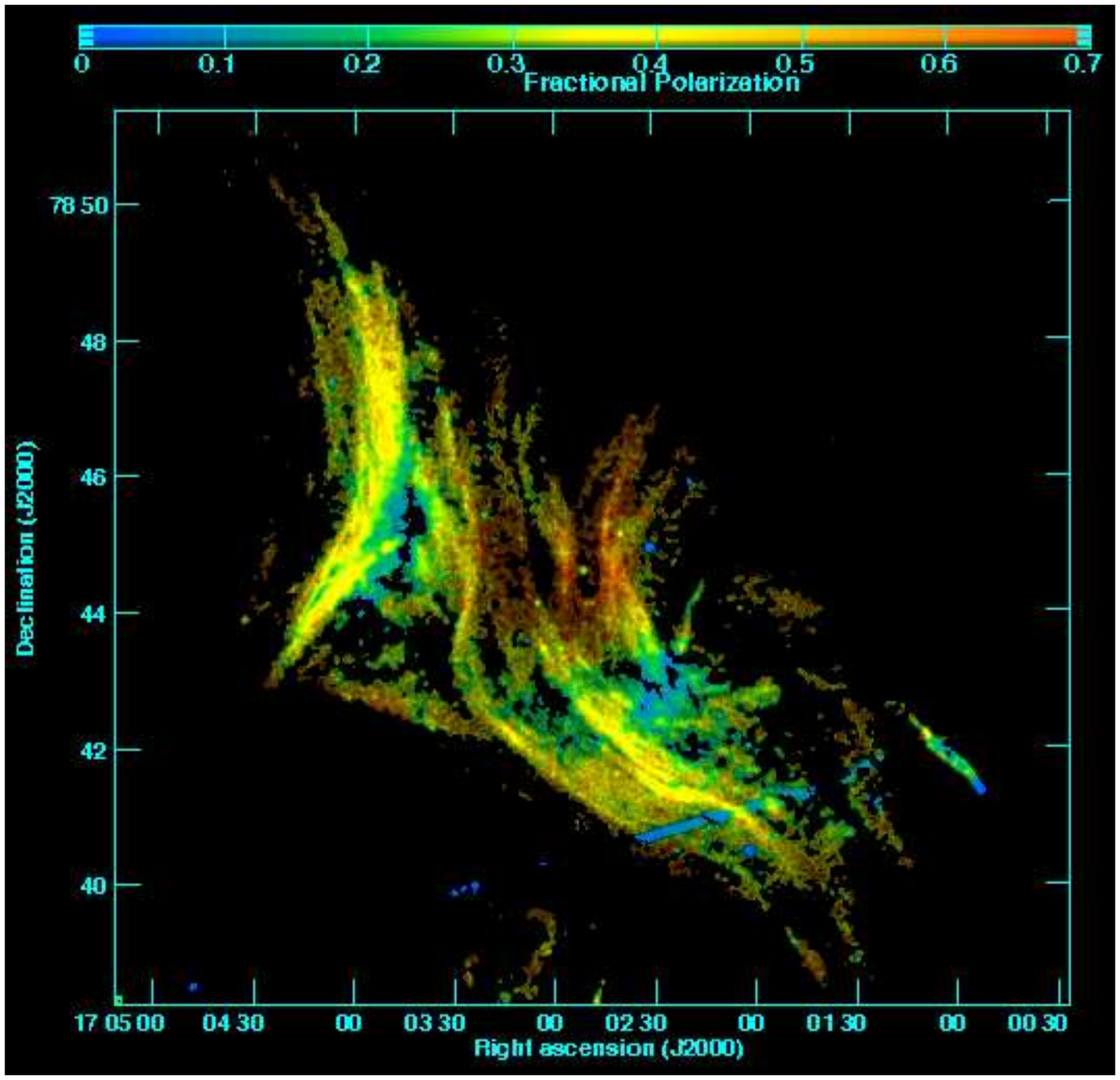}
\hspace{-0.14\columnwidth}
\includegraphics[width=1.15\columnwidth]
{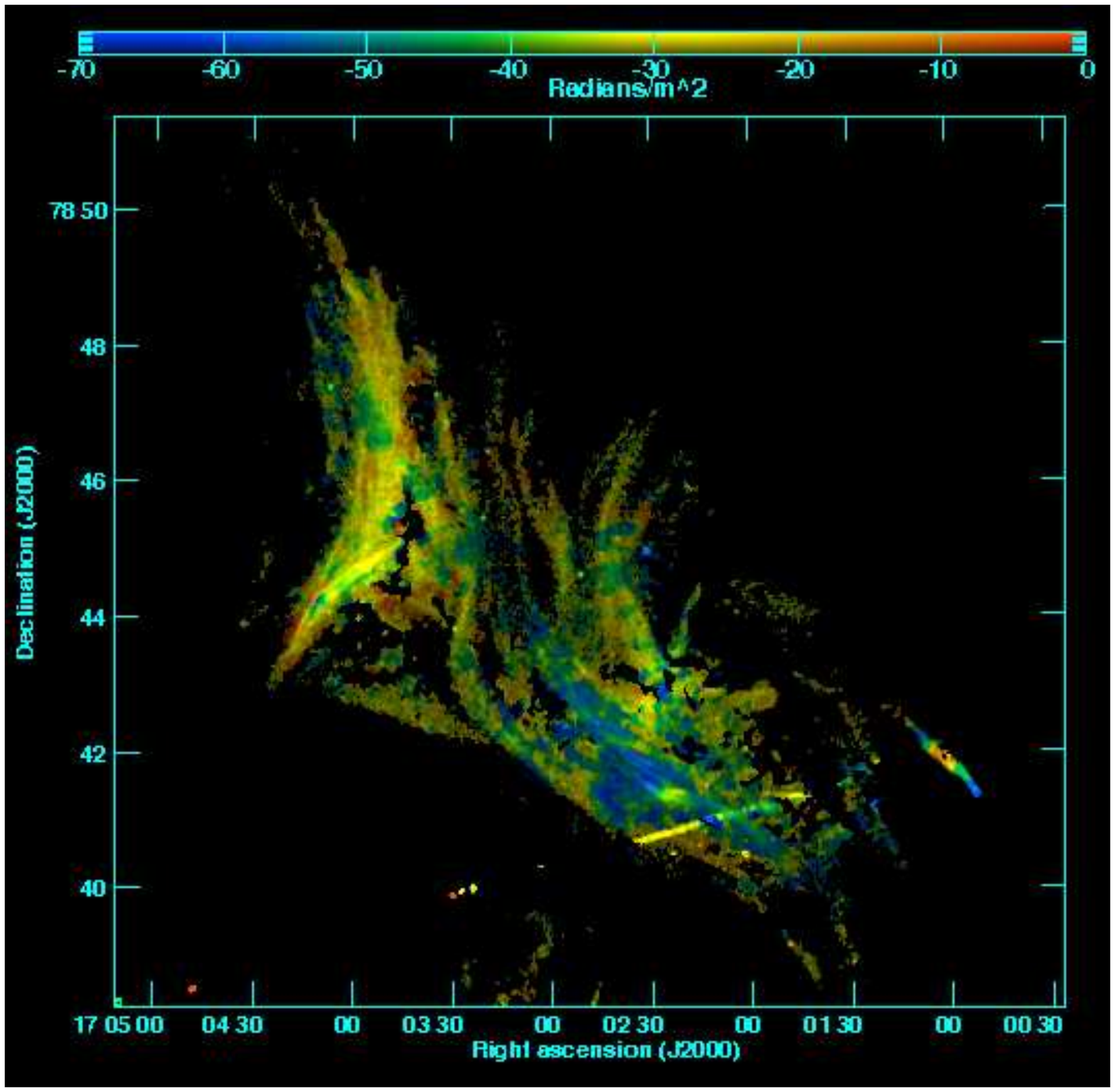}
\vspace{-1.0in}
\caption{The Large Relic:  Top left:
Large Relic at 3\arcsec\ resolution. The wedge at the top of the figure shows the intensities from
0 to 90$\mu$Jy/beam. Top right:
Relic true color image at 3\arcsec\ resolution for I, 6\arcsec\ resolution for  
the spectral index, 
$\alpha$. Lower left:
Large Relic  Intensity I-pol at 3\arcsec\ resolution; color: Fractional
  Polarization, 6\arcsec\ resolution. Lower right:
Large Relic Intensity: I-pol, 3\arcsec\ resolution; color: RM max, 6\arcsec\ resolution.
\label{ri3s}} 
\end{figure*}

\begin{figure}
\vspace{-0.9in}
\includegraphics[width=1.03\columnwidth]{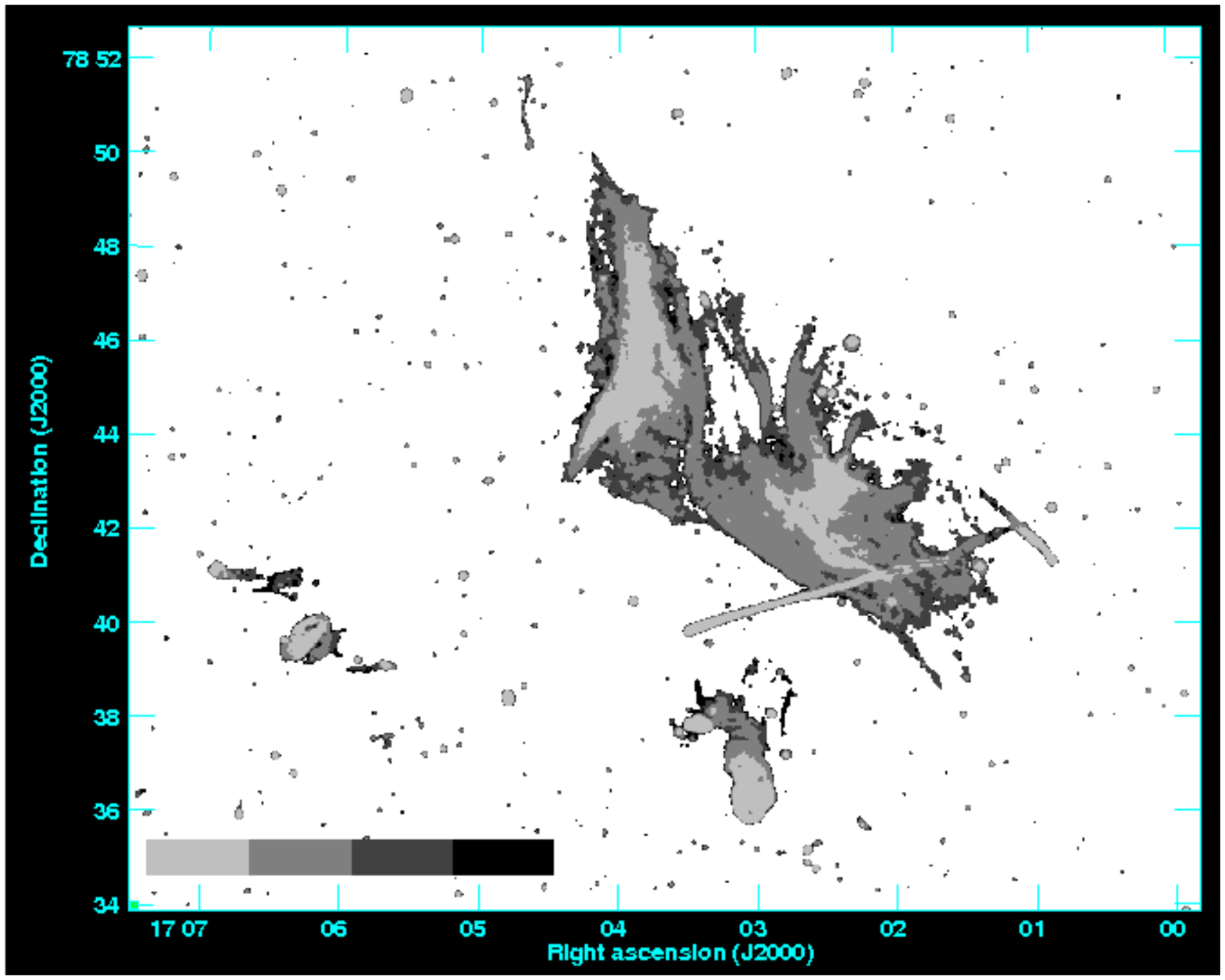}
\vspace{-0.9in}
\caption{Error in spectral index for 6\arcsec\ resolution for the area
  containing main cluster sources. Errors are given in ranges of
  $0.0-0.1, 0.1-0.2, 0.2-0.3$ and $0.3-0.4$ as shown in grey-scale
  wedge in the bottom-left of the figure.
\label{6ea}} 
\end{figure}

In this section we will discuss the brighter, extended individual
sources and source complexes. A later paper will discuss the
population of 61 radio emitting cluster members found so far on the
total intensity images, the most sensitive of which has 3\arcsec\
resolution and an rms noise near its
center of $3.7$$\mu$Jy. Other papers will also
discuss in more detail most or all of the sources in this section but the initial findings are summarized
here. In figure~\ref{bw6l}, we display a smaller field containing all the cluster features
we discuss below. The traditional letter designation is shown for the brighter individual
sources \citep[e.g.,][]{m03}. What we will call the  ``Large Relic'' is the large complex
structure which dominates the upper part of figure~\ref{bw6l}.

\subsubsection{The Large Relic}

Ever since the first detailed Westerbork images of Abell 2256 were produced \citep{b79},
the diffuse radio structures north of the cluster center have been a puzzle for 
astronomers. The most detailed study and the current standard model for the Large
Relic was produced by \citet{c06}. Their argument is that the Large Relic is due to
an outward moving shock, resulting from a cluster-cluster merger, which is located on the
near side of Abell 2256. The new wide-band VLA images allow us to revisit their
conclusions.

The radio emission from the Large Relic has been called
filamentary \citep{c06,b08} but what that means depends strongly on
resolution.
Generally, observations agree that the Large Relic does not have
an extremely steep spectrum, e.g. spectral index, 
$\alpha=-0.81$,($S\propto \nu^{\alpha}$) from 63 to 1369 MHz
\citep{v12a}, although \citet{c06} find a mean spectral index from
1369 to 1703 MHz of $-1.2$. A gradient toward flatter spectral
indices is also reported from southeast to
northwest \citep{c06,k10}. \citet{c06} report fractional polarizations up to 0.45 and
an almost uniform RM near $-44$ rad m$^{-2}$.

Our new observations show much more detail. 
In figure~\ref{ri3s} top left, we show the new 3\arcsec\ resolution
total intensity image. The Large
Relic is seen to be dominated by a complex filamentary structure. 
With the 3\arcsec\ resolution we can  not only see the filamentary web
but can also resolve most or all of the individual filaments. The shape of the
individual filaments is clearly correlated locally with other nearby filaments. The
structure hints at being made up of at least one large vortex on the northwest side 
of the Large Relic, and possibly another on the eastern edge. 
Our highest 1-2 GHz resolution image with a clean
beam of $2.15$\arcsec $\times 1.46$\arcsec\ pa$=92^{\circ}$ allows us
to resolve the widths of all the filaments. The smallest transverse
widths are $\sim 5$\arcsec\ in projection. The long filament with a center
near  17 \ffh 03 \ffm 25.9 \ffs, 78\arcdeg43\arcmin 02\arcsec\ appears
to be a twisted ribbon $\sim 5$\arcsec\ in width, which implies the Large Relic is at least 25 kpc
thick. 

\begin{figure*}
\vspace{-0.4in}
\includegraphics[width=1.0\columnwidth,trim=0.0cm 0.0cm 0.0cm 0.0cm,clip=True]
{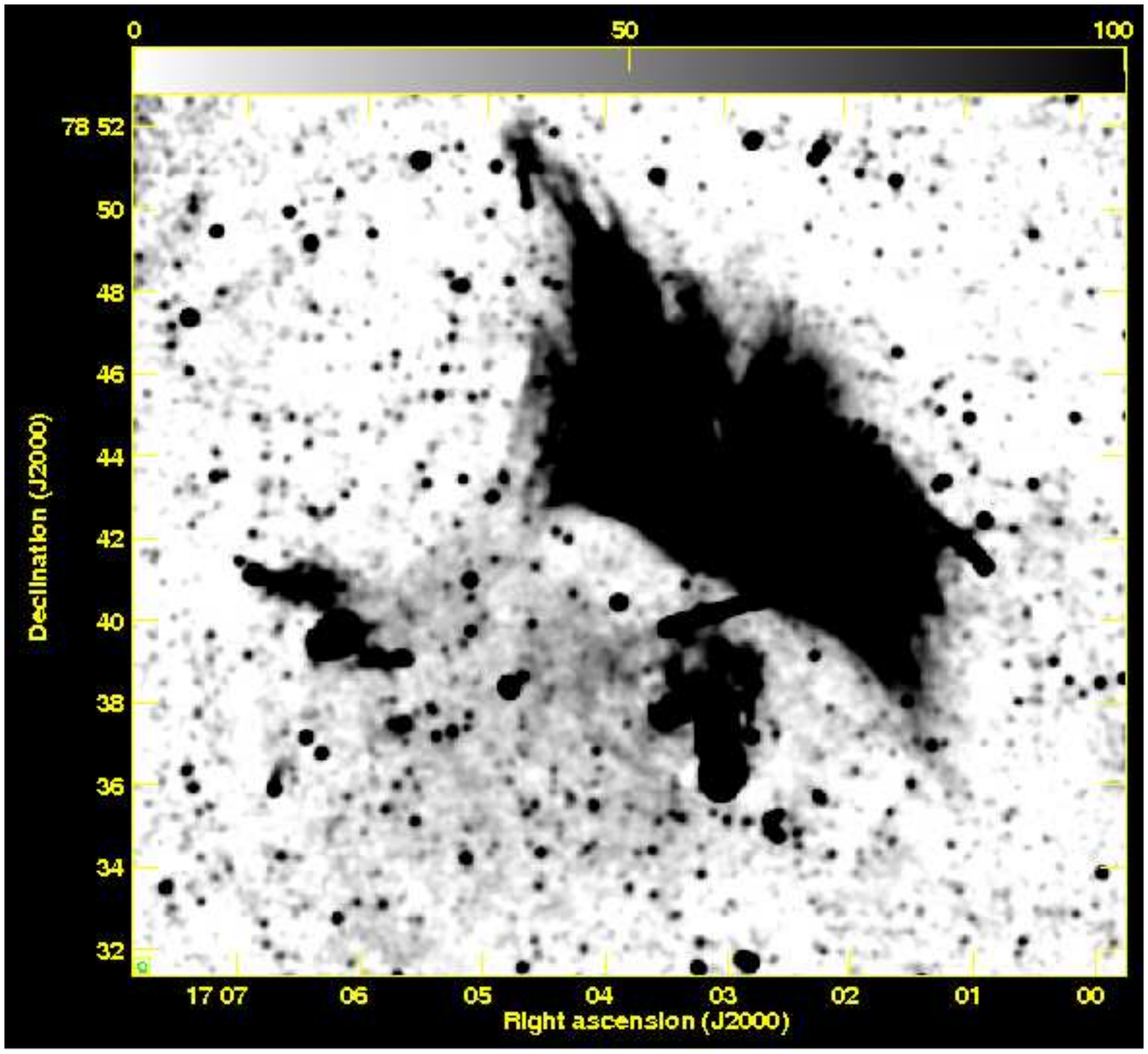}
\hspace{0.01\columnwidth}
\includegraphics[width=1.04\columnwidth,trim=1.0cm 1.5cm 1.0cm 3.0cm,clip=False]
{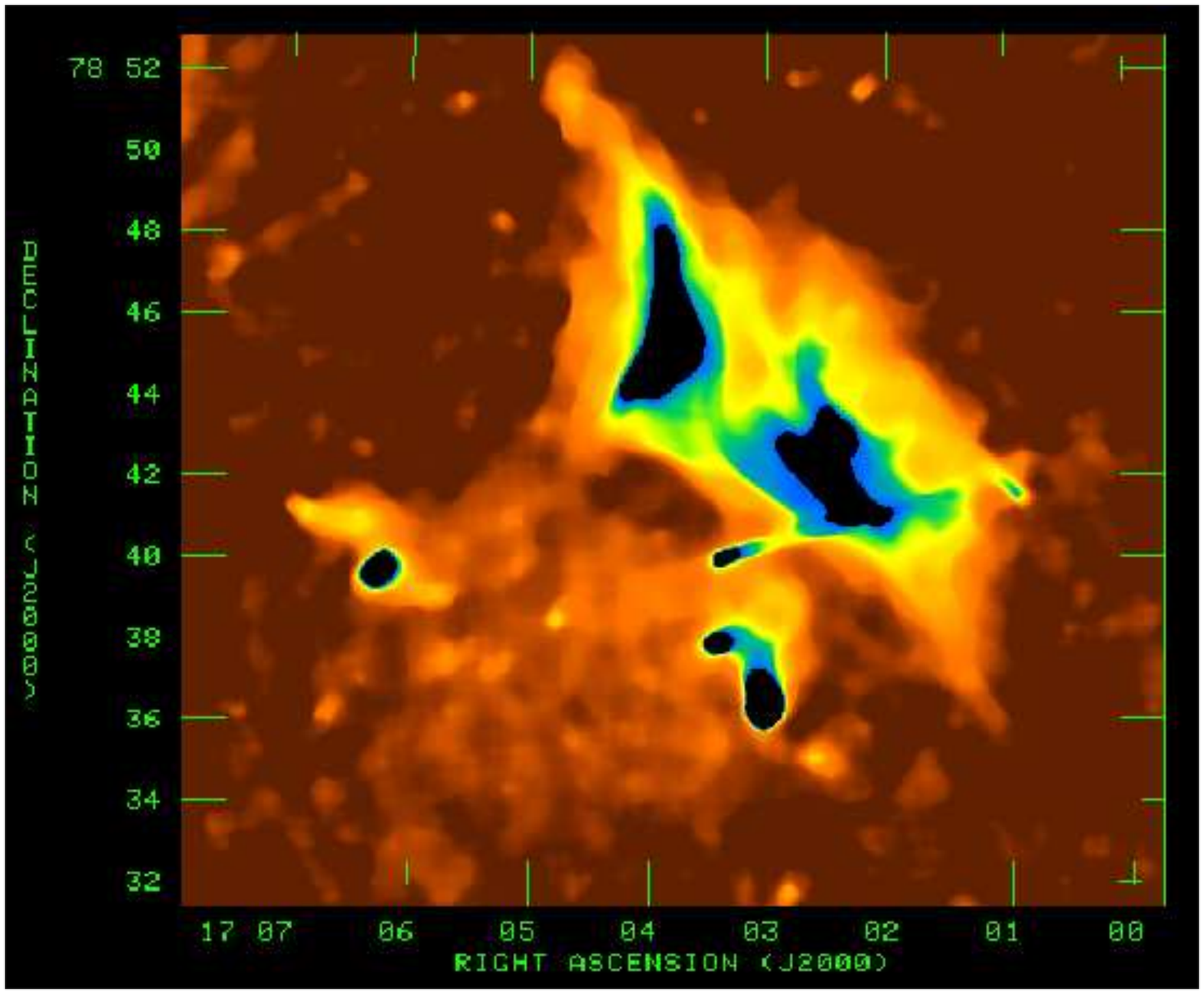}
\vspace{-0.65in}
\caption{Large Scale Structure.  Left:
  Saturated 12\arcsec\ resolution image showing ``parallelogram'' outer 4
  sharp edges of the Large Relic.
Right: Deep TVHUE Median Window Filter Lowpass image. This image shows
  the connection between  Large Relic and the Halo near the F complex.
\label{ls}} 
\end{figure*}

In figure~\ref{ri3s} top right, we show a ``true color'' image of the
Large Relic using the L-band
($1-2$ GHz) spectral
index image at 6\arcsec\ resolution to color the 3\arcsec\ total
intensity image. The overall intensity weighted spectral
index, excluding bright confusing sources and source C,  we measure to
be  $-0.94$. However, the Large
Relic is made up of a complex of different values of $\alpha$, typically
varying from about $-0.6$ to $-1.4$.\footnote{The fine scale color
  structure in this image is at least partly due to the noise in the spectral index at
  6\arcsec\ resolution. However, one can see the general trends better at this
  higher resolution than in figure~\ref{fourway_1}.} These spectral indices
are flatter than those reported by \citet{c06} and closer to the lower
frequency measurements. One can see the spectral index gradient reported by
\citep{c06,k10}; however, the more striking correlation is with the most dominant bright
filamentary structures, which have spectral indices near $-0.8$. The lower brightness
regions generally have the steepest spectra. The ends of the 
filaments on the northern side have the flattest
spectra but there are also very steep regions in the northwest and on
eastern edge as well. We show in figure~\ref{6ea}
the estimated errors in the
6\arcsec\ resolution spectral index image. 

We display in figure~\ref{ri3s} bottom left a color image of the same 3\arcsec\ total intensity 
image but in 
this image the color is the fractional polarization from the 6\arcsec\ resolution AFARS 
amplitude image divided by the 6\arcsec\ resolution total intensity image. The bright total
intensity filaments show a high fractional polarization, typically
$0.3-0.4$.  These
levels confirm the generally ordered underlying magnetic fields in the filaments that
one would guess from the total intensity image.
The filaments in the north-central part of the
Large Relic stand out as having having even higher fractional
polarization, up to 0.7. These same
filaments show the flattest radio spectra. On the other hand, there are
regions, especially in the northwest, which show signs of disorder with $\ls 0.2$ fractional
polarization. 

Thus, when the Large Relic is observed in more detail, the simple gradient from north to
south in spectral index does not dominate the picture.
Perhaps our new, more detailed images suggest that the most recent particle acceleration is associated
with the flat spectrum filaments with well-ordered fields on the north side of the relic.

In figure~\ref{ri3s} bottom right, we display a color figure of the
peak RM from AFARS. The
total intensity and field of view are the same as for the previous two figures but this time
the colors are the RMs at the peak polarized amplitude. This image, as well as
figure~\ref{ri3s} bottom left, was produced using a narrower search range
in RM ($-70$ to 0 rad m$^{-2}$) and a lower $4\sigma$ cutoff (10$\mu$Jy/beam)
after correcting for noise bias. The higher resolution and smaller
range of RM allows the lower cutoff than in figure~\ref{fourway_1}. 

This RM  image shows a 
pattern in the Large Relic not clearly correlated with the total intensity. 
 There are relatively large, coherent patches of Rotation Measure  up to a few arcminutes in
size, typically ranging from about the galactic foreground value, $-25$ rad m$^{-2}$
down to about  $-70$ rad m$^{-2}$. These patches show much larger deviations from the mean than
is consistent with studies of either Galactic RM fluctuations \citep{s09} or the local Galactic 
RM gradient across the Abell 2256, $\sim 10$ rad m$^{-2}$ per degree \citep{b08}.

 Clearly, the RM is not almost constant across the Large Relic, as appeared to be the case at the
S/N and resolution of \citet{c06}. The RM values are not greatly different than those
seen from other cluster members in figure~\ref{fourway_1} bottom right. 
Whether the  more negative RM values
are from the cluster or are local to the relic is unclear from these data. A more detailed
analysis of the polarization associated with the Large Relic will be
made in our subsequent paper. 
 
Another hint to the nature of the Large Relic is shown in the high contrast total intensity
image in figure~\ref{ls} left.  One can see that the four sides of the Large Relic have almost
sharp boundaries in a shape like a parallelogram. Some physical process must be creating
these sharp boundaries. 

Certainly the \citet{c06} model for the Large Relic,
 as an outgoing shock on the near side of the cluster, is not ruled out by these results. 
However, some of the arguments  for their model are weakened.  The X-ray images do not, 
as yet, show any  evidence of a shock coinciding with the Large Relic \citep{s02,bm,k10}.
 In addition, the \citet{c06} picture relied on the low level of RM dispersion they detected 
across the relic, which was much smaller than the value they expected if the relic were behind the 
cluster (based on their model of the intracluster medium). Our new
observations, however, reveal more 
significant RM variations across the relic, and also show that the RM values across the relic are 
comparable to those seen in other cluster members.  Thus the RM data
no longer require the relic to sit on 
the near side of the cluster. Perhaps other models should be considered, such as a large-scale current 
sheet between two magnetic domains.

\subsubsection{The Halo and Larger Scale Features}

Several papers report on the properties of the very diffuse emission south
of the Large Relic, which is called the Halo \citep{c06, b08, k10,
  v12a}. No polarization is detected and a steep radio
spectral index is found, $\ls -1.5$. These properties are generally
consistent with the class of objects called ``halos'' in rich clusters
\citep{f12}. Thus most papers assume Abell 2256 has both a halo and
one or more relics which are generally considered separately. 

Low frequency results have recently been reported from Westerbork \citep{b08,v09}, 
the GMRT \citep{k10}, and from LOFAR \citep{v12a}.
The  observations reported here were not designed to focus on the very diffuse features which could
be studied in detail with the upgraded VLA either with more integration time in the 
D-configuration or perhaps lower frequency data with the new Lowband system.
 However, our combination of resolution
and sensitivity shows one interesting large scale, low surface
brightness feature. 
In  figure~\ref{ls} right we show a median window
filtered image of our 12\arcsec\ resolution image. This image shows a connection between the
Large Relic and the Halo (the diffuse feature south of the large relic) along a curved 
arc on the eastern boundary of both features. The arc also connects the Large Relic and the
Halo near complex F which we will discuss next. It appears as if the
eastern boundaries of the Halo, the Large Relic, their connection and
possibly the F complex have some common cause, although no related
structure can be seen in existing X-ray images. 

The large scale structure of Abell 2256 is hard to understand because
of the fundamental problem of translating a two dimensional image to
three dimensions. The Large Relic has been suggested to be a
relatively thin structure in the plane of the sky which we see
projected onto the X-ray, supposedly, three dimensional
structure. However, the Large Relic and the large, curved arc are seen
in two dimensions near an outer 
the edge of the bright {\it Chandra} X-ray emission region, see
figure~\ref{BX}.  So either the Large Relic and the arc in three
dimensions are bounded by the dense cluster cluster gas, or they lie well outside of the
central cluster region and are seen in projection by
chance against the X-rays. The Large Relic itself has structures on
the eastern and north-western side which could be interpreted as being
vortices seen in projection. This could suggest that the Large Relic
is not a thin sheet but has some depth along the line-of-sight. In any
case the relic and the arc are outside of the region of high X-ray
S/N and thus we know little about the structure of the gas
in these regions.

The comparison with other relics is difficult. Most ``relics'' are
elongated, ``fuzzy'' patches without an obvious optical identification
associated with a relatively distant cluster of galaxies
\citep[e.g.,][]{f12}. A few sources, characterized as ``round relics''
by \citet{f12} could be similar to the structure in Abell 2256 but
the size and radio luminosity of most of these sources are very
different from our case. One such source associated with Abell 1664 is very similar
\citep{g01} and a few others with better radio imaging might
be in the Abell 2256 class.  However, the current radio observations
which have been used to search for relics and halos are mostly
too low in resolution to distinguish sources like the Large Relic
from halos and other types of diffuse emission. Thus the
class of Abell 2256-like relics remains to be cataloged. Clearly
such diffuse sources with similar sizes in two dimensions on the sky
and relatively high surface brightness are unusual.

\subsection{Radio Complexes and Individual Sources}

\subsubsection{Steep-Spectrum Complex F}

On the eastern edge of the cluster radio emission is a complex of
sources: Complex F. All the previous observations of Abell 2256 have
discussed this source since it is fairly well separated from the other
radio structure and shows a a very steep spectrum. \citet{r94} report
the shell structure of F2 but do not classify F3 as a Narrow Angle
Tailed radio galaxy (NAT, as defined by \citet{or}) since they
assume that the whole complex is an unusual ``Z-shaped'' single
source. For F2, \citet{b08}, \citet{k10}, and \citet{v12a} report
spectral indices over the range
from 63 to 350 MHz $\sim -1.2$ and between 350 and 1450 MHz $\sim
-1.8$. \citet{c06} measure $-2.5\pm 0.2$ between 1369 and 1703 MHz
and also find fractional polarization $<0.02$. 
 
\begin{figure}
\vspace{-0.7in}
\includegraphics[width=1.03\columnwidth,angle=0,trim=0cm 0cm 1cm
0.0cm,clip=True]
{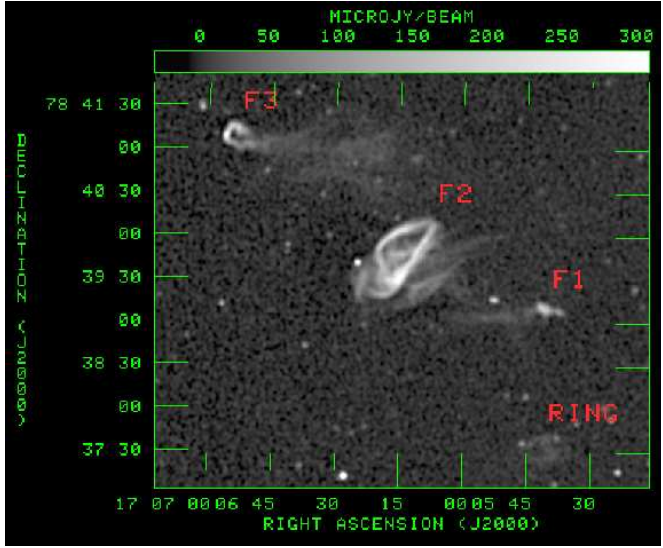}
\vspace{-1.00in}
\caption{Large F region including ring galaxy at 3\arcsec\ resolution.
\label{Flarge}} 
\end{figure}

\begin{figure}
\vspace{-0.6in}
\includegraphics[width=1.03\columnwidth,angle=0]{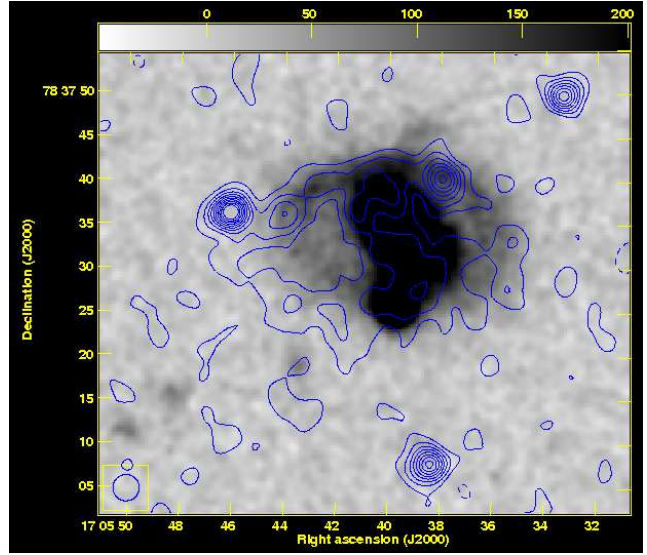}
\vspace{-0.85in}
\caption{Radio/Optical ring galaxy overlay. Grey scale is R SDSS image
  convolved with a   1\arcsec\ FWHM circular Gaussian. Contours are
  1.5e-5$\times(-1,1,2,3,4,5)$ Jy/beam. Radio resolution is 6\arcsec.
\label{roring}} 
\end{figure}

In figure~\ref{Flarge} we show a grey-scale image of four radio sources in this region
at 3\arcsec\ resolution with labels. Source F3 is a clear NAT associated
with a cluster galaxy at its far eastern edge. F2 and F1 do not have clear optical/NIR
identifications. The source labeled RING is an optical ring galaxy
\citep[]{sdss}[hereafter, SDSS] with associated radio
emission (figure~\ref{roring}) which may not have anything to do with F1, F2 or F3 
but is located in the same general region.

\begin{figure}
\vspace{-1.0in}
\includegraphics[width=1.08\columnwidth,angle=0,trim=0.0cm 0cm 0.0cm
  0.0cm,clip=True]
{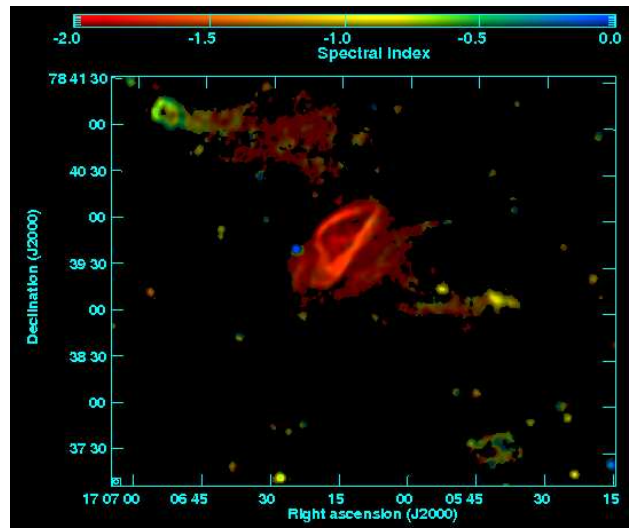}
\vspace{-1.05in}
\caption{True Color image of region F, total intensity at 3\arcsec\
  resolution, spectral index.
at 6\arcsec\ resolution.  Individual sources F3, F2, F1 and the Ring galaxy 
are labeled in Figure~\ref{Flarge}.
\label{Falpha}} 
\end{figure}

In figure~\ref{Falpha} we show a true color image of the complex showing that all three
components have very steep  spectral indices. Our image shows the
spectral index structure for the entire complex between 1 and 2
GHz. For F2 we see typical values in the bright part of the ring
of $\sim -1.8$ and the intensity weighted spectral index for F2 is
$-1.85$. Thus we do not find that F2 is steepening in spectral index
between 1 and 2 GHz in comparison with the values found at lower
frequencies. For F3 we see spectral
indices near the parent galaxy $\sim -0.7$ increasing to more
uncertain values near $-1.5$ down the tail. For F1 we find values that
range from $\sim -1.4$ to $\sim -0.8$ near its western end. While the connection of F1, F2 and F3
is unclear, it is very unusual to find three such large, unusual morphology, steep 
radio spectrum sources in so small an area on the sky. Source F3 is a fairly normal
NAT. F2 is very unusual. It has no good candidates for an optical identification but does
have an inverted spectrum point (393$\mu$Jy, 17 \ffh 6 \ffm 24.97 \ffs,
78\arcdeg 39\arcmin 41.0\arcsec) with a spectral index $\sim +0.6$ in the 1-2
GHz band, seen as a blue dot in figure~\ref{Falpha}, which appears to
be attached to one of F2's red filaments. No optical object is near the blue dot in SDSS
images, but there is a WISE \citep{w10} 3.3$\mu$m and 4.6$\mu$m detection $\sim
0.5$\arcsec\ from this position. F3 looks like it might be part of
a normal tailed radio galaxy but there is no optical identification to the limit of the SDSS
coincident with the radio emission. Unlike the lower spatial
resolution results of \citet{c06}, at
6\arcsec\ resolution, around
the bright ring we see fractional polarizations ranging from 
 $<0.02$ to $0.15$. For approximately 10 independent
beams showing strong polarization, the RMs vary
from $\sim 0$ to $-300$ rad m$^{-2}$. 
Even larger values of abs(RM) are likely present, but a more
thorough analysis of the errors is required to establish their
reality. Thus at our higher resolution F2 is polarized and has a
range of RM values which could have depolarized the source at the
lower resolution of \citet{c06}. 

 Except for F3, which clearly has an optical identification
  with an Abell 2256 member galaxy, the rest
of F could be one or more background sources seen by chance in the cluster field. The existence
of the  point source with an inverted radio spectrum associated with a WISE source also could be a 
random background source
superimposed on F. The lack of an SDSS counterpart to this point source certainly is consistent with
a dusty, perhaps high redshift, galaxy.  F1 and F2  could be
associated with this object. However, given the current data,
the existence of such an unusual morphology, very steep spectrum, large background source very close to a 
NAT radio galaxy which also has an unusually steep spectrum radio tail seems like too big a 
coincidence to call F2/F1 a background source. It also seems possible that the WISE source
is a very dusty galaxy in Abell 2256. We will have to hope that future observations make the situation
clearer.

Perhaps the object most similar to F  which has been discussed extensively is the NGC1265
complex in the Perseus cluster. That galaxy has a NAT radio source with an apparently 
attached, steep-spectrum ring \citep{sb}. \citet{pj} modeled the ring as a bubble of gas passing 
through a cluster shock. The F3/F2 complex seen from a slightly different angle might
be a similar structure. Since the F complex lies near the apparent
boundary with the large-scale radio arc (figure~\ref{ls})
these sources could have a similar origin. The fact that the F-complex
is a second example of structure
similar in morphology to NGC1265 and its adjacent ring may suggest a
more direct connection between  mass loss from the radio galaxy and the process
responsible for the ring than suggested in the ``radio phoenix'' model of
\citet{pj}.  

An interesting, related question is whether some or all of the
F-complex should be considered a ``relic'' since it has a very steep
radio spectrum, is polarized, and is an elongated structure on the outskirts
of the cluster.

\subsubsection{Central A/B Complex}

\begin{figure}
\vspace{-0.45in}
\includegraphics[width=1.1\columnwidth,angle=0,trim=2cm 0cm 1cm
3.0cm,clip=False]
{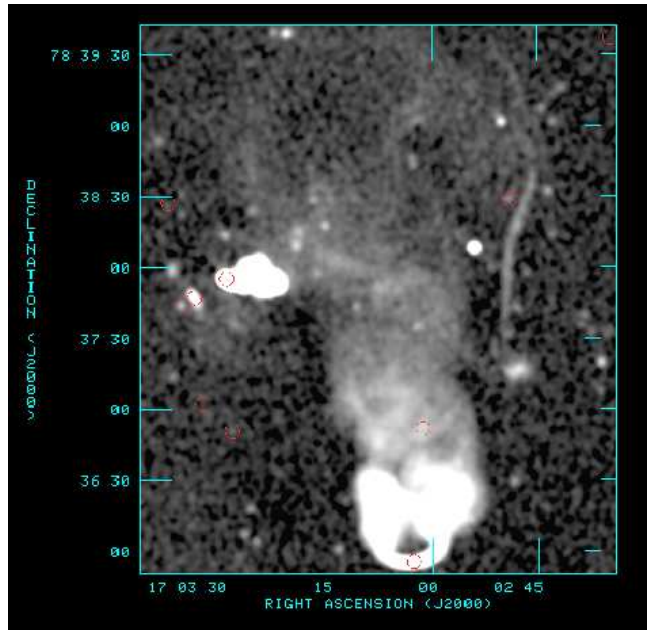}
\vspace{-1.25in}
\caption{A+B region 3\arcsec\ resolution. Red circles mark locations of
  cluster members with measured redshifts.  Individual sources A, B and the Line are
labeled in Figure~\ref{bw6l}.
\label{ABhr}} 
\end{figure}

\begin{figure}
\vspace{-1.1in}
\includegraphics[width=1.03\columnwidth,angle=0]{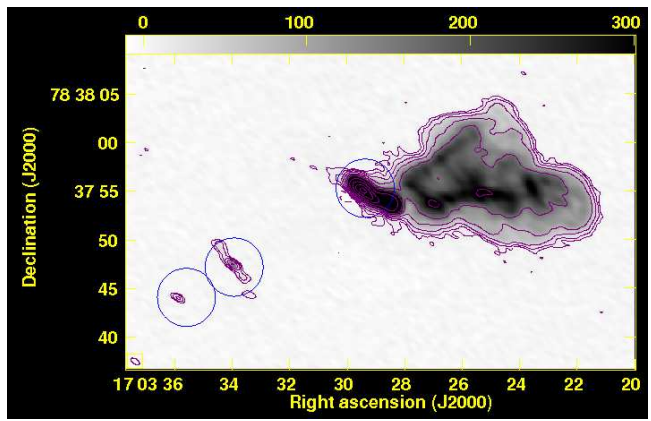}
\vspace{-1.3in}
\caption{Source A: Grey scale with contours 
at S-band. Clean beam=$1.07$\arcsec$\times0.51$\arcsec pa=57$^{\circ}$. 
 Contours are $10\times(-1,1,2,4,8,16,32,64,128,256)\mu$Jy/beam. The
 intensity wedge at the top is in units of $\mu$Jy/beam. The blue
 circles are 3\arcsec\ in radius and show the locations of three galaxies in the NGC6331 triple system.
\label{AS}} 
\end{figure}

\begin{figure*}
\vspace{-3in}
\includegraphics[width=2.1\columnwidth]{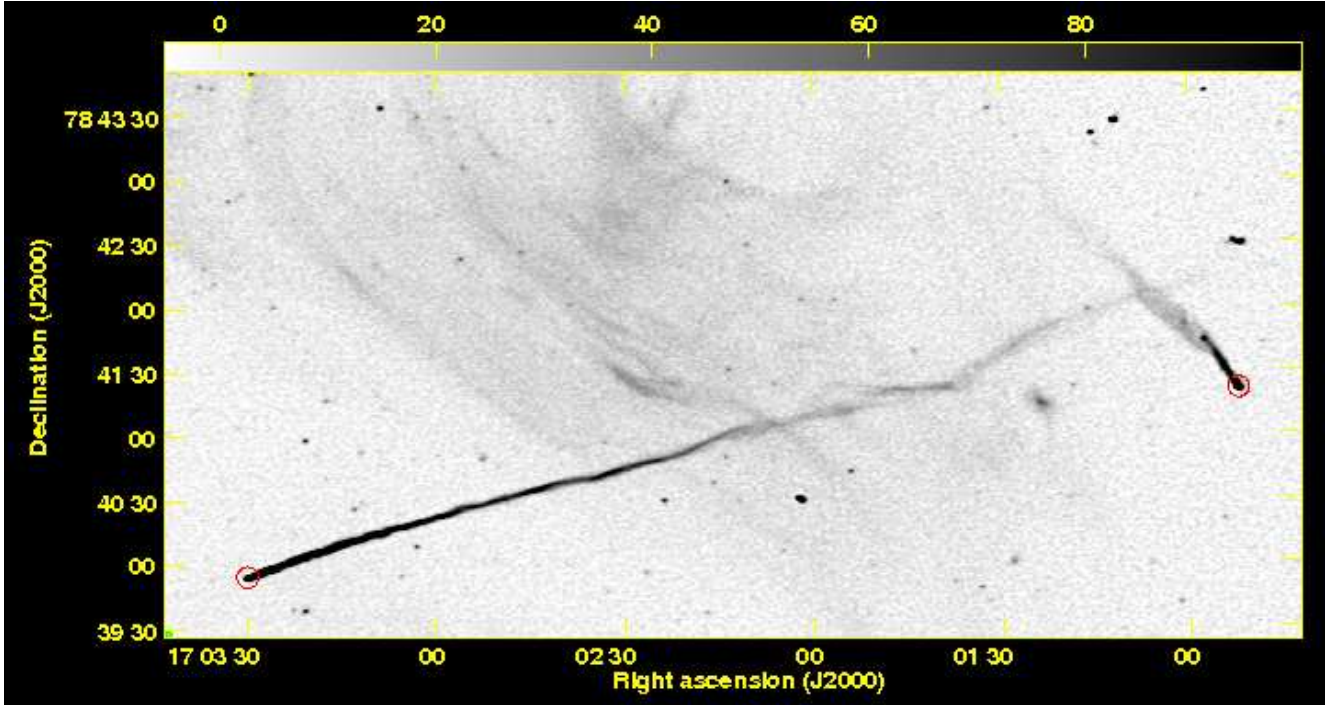}
\vspace{-2.62in}
\caption{Sources C and I at L-band full resolution with optical IDs
  (red circles). 
Clean beam=$2.15$\arcsec$\times1.46$\arcsec pa=92$^{\circ}$.  The green
  circle showing the beam is in the lower left corner of the image.
\label{CIL}} 
\end{figure*}

\begin{figure}
\vspace{-0.83in}
\includegraphics[width=0.96\columnwidth,angle=0,trim=0.5cm 0.0cm 1.5cm
  3.0cm,clip=False]
{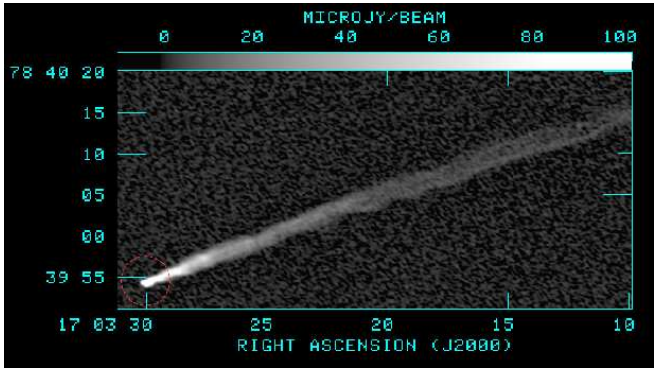}
\vspace{-1.3in}
\caption{Source C head in 4-6 MHz band with 3\arcsec\ radius red circle on
  optical ID. Clean beam=$0.57$\arcsec$\times 0.34$\arcsec
  pa=64$^{\circ}$. 
\label{CSr}} 
\end{figure}

\begin{figure}
\vspace{-0.2in}
\includegraphics[width=1.0\columnwidth,angle=0,trim=2.0cm 0.5cm
1.0cm 5.0cm,clip=False]
{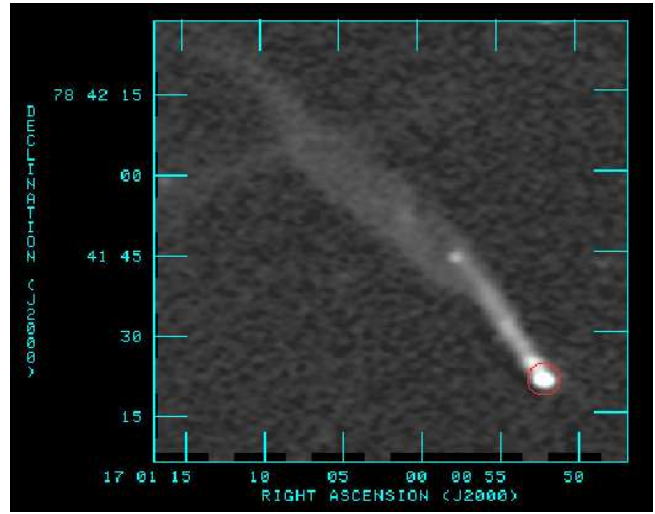}
\vspace{-1.25in}
\caption{Source I: L-band Clean beam=$2.15$\arcsec$\times1.46$\arcsec
  pa=92$^{\circ}$. The red circle shows the location of the center of
  the optical identification. 
\label{Ihr}} 
\end{figure}

Near the center of the field below the Large Relic is the A/B Source Complex. 
Figure~\ref{ABhr} is a grey-scale image of the complex including a large
NAT (source B) running from the 
bottom of the figure at least half way to the top. Source A is the small, very bright, 
extended blob to the east of source B. Earlier work at
1400 MHz with the VLA \citep{r94,m03} showed the 
horseshoe-shaped morphology of B, characteristic of a NAT. Recent images at
frequencies below 1 GHz \citep{b08, v09, k10, v12a}, show that there
is complicated, steep-spectrum structure along the northern extension of B.

Our new images show more details. At the top of source B (or just
above it) there is perhaps
a faint ring and/or some other fine-scale structure covering
the upper half of the image. To the west of B there
also are one bright ``Line'' and perhaps another fainter line above
it. In the upper left panel of figure~\ref{fourway_1}, one can see that
source B shows a steepening spectral index 
like most NATs \citep[e.g.,][]{sb,l04}  as well as  a very steep spectrum region
corresponding to what is seen on the lower resolution, lower frequency
images \citep{b08,v12a}.  It is
not clear whether this region is part of B, part of a separate ring or in some way related
to A.  The Line, not clearly seen in the lower frequency images, also has a very steep spectrum and is
$\sim 40$\% polarized.\footnote{The slightly extended blob below the Line is associated with a
galaxy which is not a cluster member.} Perhaps the Line is part of the
steep spectrum region north of B. Our new imaging raises the question of
whether all the steep spectrum structure in the northern half of the
complex is part of B or one or more distinct emission regions
projected as if they were connected with  B. 
In figure~\ref{BX}, the X-ray overlay, one can see a strong gradient in the X-ray emission
which is a cold front \citep{s02} slightly to the north of the steep spectrum radio bar
and just south of the ring. Is this radio structure somehow related to
the cold front ?

In figure~\ref{AS}, we show the S-band (2-4 GHz) A-configuration image of source A and 
friends, with contours overlaying the grey-scale. The optical identification of A is
apparently part of a triple galaxy, which as a unit should probably be
called NGC6331 based on the NGC catalog \citep{d88}, which is the only
NGC galaxy and the brightest galaxy in the cluster. The entire system
appears to be contained in a single diffuse optical halo on the  SDSS r image with a major axis
$>70$ kpc in diameter. In our image
one can see radio emission from all three components of the triple,
which have a range of radial velocities of 1728 km/s \citep{b02}, not unusual
for such a multiple system but likely indicating that the individual
components are on highly radial orbits that pass near the core of the
cluster \citep{t85a,t85b}.  The  radio galaxy A was included in the
original list of six NATs by \citet{ro}. \citet{r94} also 
describe A as a definite head-tail source (i.e. a NAT). 
However, source A has a much higher surface brightness than any other
radio galaxy in the cluster and the sharp boundaries of the
radio source seems to indicate a source more like a  bubble than a tail.
Source A appears to be small on the cluster scale; however, it
is useful to compare A with the central cluster radio source
associated with M87 \citep[e.g.,][]{o00}, a buoyant bubble blown in the
cluster gas at the center of the Virgo Cluster \citep[e.g.,][]{c01}. 
The major axis of A is $\sim 31$ kpc
in size, about half the full size of M87 radio source. Its 20cm luminosity is
$1\times 10^{24}$ W Hz$^{-1}$, about 7 times smaller than M87; however,
the brightness per unit volume is very similar. One can see a double
structure in figure~\ref{AS} on the eastern side of the source, about 
5 kpc
in size, very similar to the inner double in M87 \citep[e.g.,][]{h89}.
Thus, except for the offset structure of the bubble, the parameters of
source A seem  more similar to a central cluster bubble than a NAT.

\subsubsection{The Long Tail C and tail I}
Perhaps the second most striking feature in the cluster is the Long
Tail, source C.  VLA images of both C and I are reported by \citet{r94} and 
  \citet{m03} at 1400 MHz and at 325 MHz by \citet{r94}.  
In particular \citet{r94} considers the Long Tail C as a NAT.
  They show that the source remains straight and unresolved along much of the
  tail and that the spectral index steepens along the tail.
 They conclude it is likely not a jet since it doesn't show a
clear nuclear component and  because  most low luminosity jets are
 two-sided. They then consider the problem of how a trail produced by a galaxy
 orbit could follow the observed, straight path. They conclude that
  such an orbit is possible if the initial velocity is two or three
  times higher than the cluster velocity dispersion, the galaxy
  orbit is well outside the merging region, and the bent-back, twin jets at
  the galaxy nucleus are hidden  by their $1.2$\arcsec\ $\times
  1.4$\arcsec\ resolution.

In figure~\ref{CIL} we show the grey-scale image at the full
L-band ($1-2$ GHz) resolution.  On our images source C is
$\sim 540$ kpc in length in projection. The tail could be $\sim800$ kpc
long if the steep spectrum  features, AG+AH, are part of the tail 
\citep{v09, v12a}. Also plotted in the field is source I and the positions of the optical
identifications are shown as small red circles. Note that C wiggles and perhaps interacts
with the Large Relic but doesn't show split trails. 
However, in figure~\ref{CSr} we display the 4-6 GHz, A-configuration image of the
head of C at $\sim0.5$\arcsec\ resolution.  Here we see the bright
core and some bifurcation of the tail downstream but not near the central component. In fact
the ``trail'' emission seems more to be limb brightened and to show twisted filaments like
many radio jets, e.g. M87 \citep{O89}.
Near the core the source has only one strand and is $\ls 100$pc in diameter, based on a
fit to the 6-8 GHz image which has a resolution $\sim0.3$\arcsec\
across the inner part of the source. Thus we do
not see the bifurcation expected near the core of the twin, bent-back jets normally seen in
Narrow Angle Tail sources.  Apparently the Long Tail could be a one-sided jet. Perhaps the
other side of the twin jet has been disrupted by the ram pressure due to the motion of
the galaxy relative to the cluster IGM or perhaps the limb
brightening we see is an indication of an underlying twin tail
structure. If so, the bending of the twin tails must take place 
very deep in the galaxy core where one might expect little impact from
the galaxy's motion through the IGM, unless the galaxy has been almost
completely stripped.  

\begin{figure}
\vspace{-0.7in}
\includegraphics[width=1.03\columnwidth]{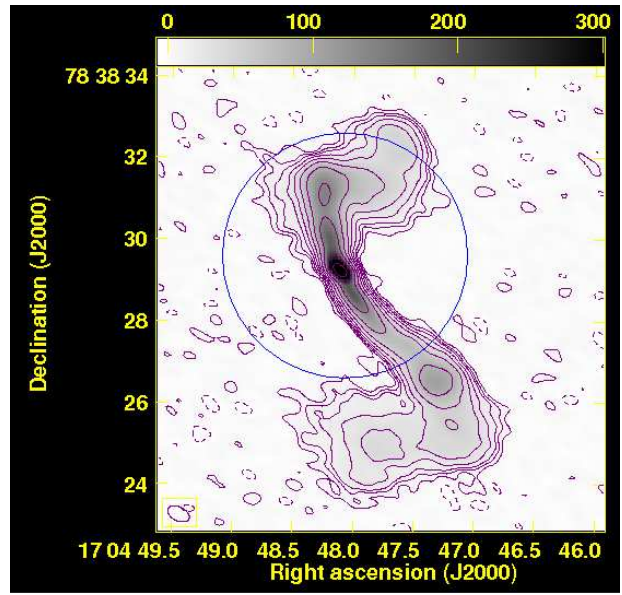}
\vspace{-0.8in}
\caption{Source D: $4-6$ GHz image with circle for optical 
 Contours are $4\times(1,2,3,5,8,13,21,34,55,89,144)$
 $\mu$Jy/beam. The intensity wedge at the top is units
  of $\mu$Jy/beam. Clean beam=$0.57$\arcsec$\times 0.34$\arcsec
  pa=64$^{\circ}$. The 3\arcsec\ blue circle shows the location of the
 optical galaxy. 
\label{Dhro}} 
\end{figure}

In figure~\ref{Ihr}, we show the S-band image of source I. Close to its
parent galaxy  source I also seems to be primarily
a one-sided jet. At the end of the inner part of the ``tail'' there is a ``knot'' or perhaps
a ``hot-spot'', just before the tail changes direction and becomes more of a diffuse plume
as seen in figure~\ref{CIL}. This morphology certainly resembles a jet
ending in a hot-spot. The direction of the more diffuse structure
beyond the knot then changes, suggesting that the relative
velocity vector of the external medium is not along the direction of
the inner structure. Perhaps not all NATs cited in the literature are
twin jets bent back by ram-pressure. In
these two cases if the direction of the ``trail'' is related to the
motion of the galaxy relative to the external medium, the second jet
has not remained stable.

Sources C and I raise the interesting question of whether single thin
structures that appear similar to one-sided jets can actually be tails
swept back by the relative motion through the ICM, similar to the
twin-tailed NATs.  In this case, the question remains whether only a
single jet was launched from the AGN, or whether a second jet was
launched, but quickly disrupted.   VLBI observations of apparently
single tails might resolve this issue.

\subsubsection{Source D}

\citet{r94} show an image of source D and the location of its parent
galaxy suggesting it is not located near the galaxy center.
In figure~\ref{Dhro} we show a grey-scale plus contour plot of source D, made at S-band in the
A-configuration. The 3\arcsec\ ($\sim 3$ kpc) radius blue circle marks the
  centroid of the optical galaxy based on the SDSS image. This FRI is
  entirely contained within the isophotes of the optical galaxy. The
  galaxy is almost centered on the core of the radio source emission.

\section{Discussion}

Abell 2256 is perhaps the most striking collection of radio
structures in a rich cluster. Why is
that? Abell 2256 is a fairly rich cluster, Abell richness class 2, but clearly less
rich than Coma, which is almost richness class 3. Coma possesses a radio halo and several
radio galaxies but seems dull in appearance compared to Abell 2256. Perhaps the objective
reason for the appearance of Abell 2256 is the surface density of complex radio emission. But
why is there so much going on?

In this paper we have presented more detailed observations of many of the sources than
have previously been published. Many show unusual properties. Certainly the relatively
flat-spectrum Large Relic takes up the most surface area. While there are other sources
called relics which are as luminous and have as large a linear extent in one dimension,
we are unaware of such a large two-dimensional area of emission with an average spectral 
index at 20cm near $-0.9$. Certainly there must be other such sources but apparently not
so nearby. 
There are also several very steep spectrum extended regions of radio emission.
Most prominent are the F-complex, the Line, and the ridge at the top of Source B. Except for the
NAT, F3, none of these sources have obvious optical identifications. Along with the Large
Relic these sources seem to be non-AGN ``cluster sources''.  
The Long Tail, source C, which could be a very long, one-sided jet, is also unusual if not 
unique. 

\subsection{Abell 2256, an off-axis, mid-merger event ?}

What physical properties of Abell 2256 could be responsible for this collection of
unusual radio structures? Certainly the mass, X-ray luminosity and the cluster temperature 
are not unusual. However, in addition to the apparent three
groups in the galaxy velocity space, the {\it Suzaku} observations of
\citet{t11} show that  the two peaks in the X-ray image
in figure~\ref{BX} differ in radial velocity by $\sim 1500$ km s$^{-1}$
so there is strong evidence for a very active merger. \citet{vr} have recently reported a simulation of a
group-cluster merger which may provide a context for understanding
Abell 2256. This simulation shows that special, extreme conditions
occur close to the time when the group passes close to
the centroid of the larger cluster. The relative velocity of the
galaxies in the group with respect to the cluster galaxies and the
external medium reaches a maximum about three times greater than for
galaxies in a similar isolated cluster. The simulation also shows that
the ram pressure on the
galaxies, especially those in the group, increases by up to two orders
of magnitude (supposedly due to relative motion of gas and galaxies).  
The interaction of the ICM in the group with the larger
cluster transfers energy to the cluster ICM in the form of shocks, flows and
general heating. This interaction has the potential to create new
boundaries in the external gas which could be seen in radio
emission. 

Other simulations of mid-merger conditions also produce interesting
insights into possible physical conditions in Abell 2256. In
particular, \citet{t00}, \citet{r98} and \citet{rs} simulate slightly off-axis
mergers which produce complex, asymmetric conditions especially near
mid-merger. \citet{t00} and \citet{rs} both show that at times close
to mid-merger there are isolated high temperature regions which wrap around 
parts of the outer boundaries of the two X-ray intensity peaks. \citet{r98} show similar
results tuned to match the X-ray results for Abell 754 but
also show the velocity field in the gas and the location of the group
ICM relative to the X-ray brightness. This simulation shows that the
unseen, heated group gas is flowing toward the brightest X-ray peak, creating
a sharp boundary and  a weak shock. While these simulations should not
be expected to duplicate Abell 2256, they suggest that part of the Abell
2256 morphology could be due to an off-axis merger.

Based on comparing the Takizawa work with the {\it Chandra} X-ray
observations, \citet{s02} concluded that Abell 2256 is seen in
mid-merger. The detailed line-of-sight galaxy velocities also seem
consistent with that possibility.
\citet{b02} report three distinct statistical, velocity/spatial groupings in Abell 2256:
a primary cluster with 186 members, a large infalling cluster with 78 members and a group
with 30 members.  These results seem consistent with \citet{vr}. In
particular, near mid-merger the
simulation shows that the group splits into two velocity components, in addition to the remaining main cluster
distribution. Whether there are two or three mass components the overall
picture is consistent with Abell 2256 being seen at mid-merger and in
a not quite head-on merger. Also, interesting interactions
are likely going on the edges of the main X-ray emitting brightness distribution.

This leads to possible interpretations of the source C, and perhaps
source I, as
due to the much increased ram pressure on the group members and relative velocities between
the group members and the cluster gas. The simulation of \citet{vr}
shows that the ram pressure can rise above $10^{-10}$ dyne cm$^{-2}$
close to mid-merger. Their models show the pressure is high enough to completely
strip galaxies, in agreement with \citet{gg} and the more detailed stripping
simulations of \citet{rb}. 
 
The F complex lies $\sim 500$ kpc in projection from the  center of
the cluster
near a boundary shown in figure~\ref{ls} beyond the edges of the
bright X-ray emission in
figure~\ref{BX}. Perhaps this suggests that these features are related to
lower density gas flowing toward the main cluster reservoir stripping
the galaxy associated with F3. \citet{vr} also show that 
the mid-merger interaction between the group and
the cluster increases the cross-section for galaxy collisions, which
could be responsible for the ring galaxy near the F complex shown in
figure~\ref{roring}. These unusual radio galaxies may be showing us
that the increased ram pressure at mid-merger dominates the
stripping and thus the evolution of cluster galaxies, rather than the
interactions during the much longer, benign phases of a cluster's
life.

Thus while the discussion above is far from definitive, the off-axis
mid-merger hypothesis has aspects that could be
responsible for origin of the large number of unusual
features seen in Abell 2256. If this hypothesis is correct, studies of
other mid-merger clusters in detail, looking for similar phenomena,
offer a new way to investigate the evolution of clusters and the
galaxies in them. 

\subsection{The Large Relic Physics}

Radio relics on the outskirts of clusters are typically interpreted as
cluster merger shocks, which are radio-loud due to in situ particle
acceleration;  the  Large Relic in Abell 2256 has specifically been
interpreted this way \citep{c06}.  
However, this object does not fit easily into the simple picture, for
several reasons. 

\begin{enumerate}

\item  The Large Relic has an unusually large aspect ratio, being nearly
as wide as it is long.  Most other, well studied relics are long, thin structures, 
no more than $\sim 100$ kpc wide (e.g, the ``sausage'' relic \citealt{v10}, 
or the ``toothbrush'' relic, \citealt{v12t});  
similar widths are predicted by theoretical models
\citep[e.g.,][]{k12,s13}.
  
\item  The Large Relic is polarized at significant levels, which suggests
  large-scale ordering of the magnetic field in the sky plane.  
It also has large, coherent Rotation Measure patches, 
which suggests large-scale ordering of the magnetic field along the
sight line.  By comparison, merger-shock models predict the 
magnetic field will show disorder on small scales in the shock plane 
\citep[e.g.,][]{s13}; thus the relic should be polarized only when
viewed edge-on.  
 
 
\item  It is not clear that the  relatively weak merger shocks (Mach number $\sim 2-3$) expected 
from simulations can account for radio-loud structures such as the Large Relic.  Much stronger shocks 
are probably needed in order to accelerate a significant number of electrons from the thermal pre-shock 
plasma and amplify the pre-shock magnetic field to useful levels \citep[e.g.,][]{g00,bj}. 
Alternatives such as reacceleration of fossil cosmic-ray electrons \citep[e.g.,][]{p13}
are challenged by the need to have a pre-existing Mpc-scale fossil structure.  
Furthermore, no evidence for such a shock close to the large relic has yet been found in the 
X-ray data on this cluster \citep{s02,b08}.

\end{enumerate}

    These arguments do not, of course, preclude the Large Relic being an
unusual example of a merger shock.  That model is still worth
consideration, but we also want to explore alternative models.  
One such class of models would have the Large Relic being the result of 
a large-scale current sheet, sitting  at the boundary between two
magnetic domains \citep[e.g.,][]{pf}
involved in the ongoing merger (as in, for instance, the simulations 
of \citet{r98}).  We will present a more detailed study of both models 
in a subsequent paper on the Large Relic, but a few points are worth
noting here.   In a magnetized plasma environment, such as the
intracluster gas involved in an ongoing merger, current sheets arise 
naturally at plasma boundaries and also at velocity shear surfaces.   
Current sheets are well known to produce magnetic flux ropes when 
they become unstable to the tearing mode \citep[e.g.,][]{pf}, 
providing a natural explanation for the dramatic filaments in the
Large Relic.  Magnetic reconnection across current sheets also
provides an alternative source of  particle acceleration.  While this
process is not as well understood as diffusive shock acceleration, two mechanisms 
have been suggested. The electrons may gain energy by falling through 
the large-scale potential drop in the current sheet
\citep[e.g.,][]{rl,bp}.  They may also undergo first-order Fermi
  acceleration if the plasma converging on the reconnection region
  are turbulent \citep{dl,d}.  While neither of these mechanisms is as
  well-studied as is shock acceleration, both seem likely ways to make
  a merger-driven current sheet ``light up'' as a radio relic.

\section{Conclusions}

The new, high resolution, high sensitivity observations with the
upgraded VLA reveal many new details about Abell 2256.

\begin{enumerate}

\item The Large Relic is made up of a complex system of filaments
  which also show locally correlated topology. Complicated variations in spectral index, 
  $\sim -0.6$ to $-1.4$ are seen across the source, with steeper
  spectra tending to be in lower surface
  brightness, less filamented regions.  Fractional
  polarization is in the range $\ls 0.1$ to $0.7$,  and coherent patterns of
  Rotation Measure are $\sim 100$ kpc in size. Essentially all the
  filaments are $\ls 5$ kpc wide. The shapes of the
  filaments are correlated over $\sim 200$ kpc
. One apparently twisted filament
  suggests that the width of the Large Relic is at least 25
  kpc. 

\item An arc of low surface-brightness emission appears to
  connect the Large Relic with complex F and the radio halo,
  suggesting a physical connection between these features. 

\item The steep-spectrum source, F2, is a polarized, pseudo-ring structure and doesn't
  appear to be directly connected with any radio galaxy, although it
  has interesting similarities to the ring associated with NGC1265
  which is modeled as bubble that has passed through a shock
  \citep{pj}.

\item The northern half of the A/B complex shows a 
  bubble-like structure, with steep spectrum, filamentary structures. The most
  prominent is a thin ``Line'' of emission west of B which has a
  very steep radio spectrum and a fractional polarization of $\sim
  0.4$.

\item The relatively high surface brightness and sharp edges of source
  A suggest it is a bubble with more physical and morphological
  similarities to  cluster center sources like M87 than to NAT radio galaxies.  

\item The Long Tail, source C, does not show a bifurcated structure
  near the core, as one would expect for a radio trail, and is $\ls
  100$pc in diameter. The source could either be due to extreme
  stripping of the galaxy's ISM due to the cluster/group merger and/or it
  could be a one sided jet. Source I also shows a similar,
  single-strand structure near its nucleus. 

\item  The unusual radio phenomena seen in Abell 2256 could be due
  to the cluster being seen at near mid-merger of a slightly off-axis
  collision of a cluster and a group.

\item Given the lack of evidence for a strong shock near the Large
Relic which could accelerate the relativistic electrons seen as
synchrotron emission, it is worth considering other models such as
reconnection between two magnetic domains.

\end{enumerate}

\section{Acknowledgments}

We wish to thank Eric Greisen for help with the development of several
AIPS tasks and verbs used in our analysis.

Partial support for the work of LR is provided through NSF grant
AST-1211595 to the University of Minnesota. 

Funding for the SDSS and SDSS-II has been provided by the Alfred
P. Sloan Foundation, the Participating Institutions, 
the National Science Foundation, the U.S. Department of Energy, 
the National Aeronautics and Space Administration, the Japanese
Monbukagakusho, the Max Planck Society, and the Higher Education Funding Council for
England. The SDSS Web Site is http://www.sdss.org/.

  This publication makes use of data products from the Wide-field
    Infrared Survey Explorer, which is a joint project of the University of
    California, Los Angeles, and the Jet Propulsion Laboratory/California
    Institute of Technology, funded by the National Aeronautics and Space
    Administration.

\clearpage

\end{document}